\documentclass[aps,prb,preprint,groupedaddress,showpacs]{revtex4}
\usepackage{epsfig}  
\usepackage{graphicx}
\usepackage{amssymb}                               
\begin{document}

\title{In situ measurements of density fluctuations and compressibility in silica
glasses as a function of  temperature and 
thermal history}
\author{C. Levelut},  
\email{claire.Levelut@lcvn.univ-montp2.fr}
\author{A Faivre, and  R. Le Parc} 
\affiliation {Laboratoire des Collo\"{\i}des, Verres et Nanomat\'eriaux, CNRS/UMR5587,
Universit\'{e} Montpellier II, cc 69, 34095 Montpellier cedex, France}

\author{B. Champagnon}
\affiliation{ Laboratoire de Physicochimie des Mat\'eriaux Luminescents, Universit\'e Lyon 1, CNRS/UMR 5620, 69622 Villeurbanne Cedex }

\author{J.-L. Hazemann}
 
\affiliation {Laboratoire de Cristallographie, UPR 5031, 26 avenue des
 Martyrs, BP 166,
38042 Grenoble CEDEX 9, France and LGIT, CNRS/UMR 5559, Universit\'e Joseph
Fourier, 38402 Saint Martin d'H\`eres, France }

\author{J.-P. Simon}

\affiliation{Laboratoire de
Thermodynamique et Physicochimie M\'etallurgiques, UMR 5614, INP et UJF
Grenoble, BP 75, 38402 Saint Martin d'H\`eres, France}

\date{\today}

\begin{abstract}

In this paper, small-angle X-ray scattering
 measurements are used to   determine the different compressibility
contributions, as well as the isothermal compressibility, 
$\chi_T^0$, in thermal equilibrium  
   in silica glasses having different thermal histories. 
Using two different methods of analysis, in the supercooled liquid and 
in the glassy state, we obtain respectively the temperature and fictive
 temperature dependences of $\chi_T^0$. The values obtained in the glass and 
 supercooled liquid states are very close to each other.  They agree  with 
previous determinations 
of the literature. The compressibility in the glass state slightly 
decreases with increasing fictive temperature. The relaxational 
part of the compressibility is also calculated and compared to previous determinations.
 We discussed 
the small differences between the different determinations. 
\end{abstract}


\maketitle

\section{Introduction}

The nanometer range in glasses is expected to 
correspond to fluctuations of density and composition, whereas at shorter
scale the structure is very close to that of the crystalline counterpart and
at larger scale the medium can be considered as  homogeneous. Those
 fluctuations, or part of them, are  frozen-in
during the cooling process and thus they are determined by the cooling
 rate and
 the thermal history of the glass sample, which can be characterized by a 
``fictive temperature'' for a certain property (see Ref.\    ~\onlinecite{Tool}
 for more details).
 As a consequence, the fluctuations are expected to depend on
the fictive temperature which will be used in this paper to characterize
 the thermal history of each glass sample.

One possible experimental method to study electronic density fluctuations
at the nanometer scale is Small-Angle X-ray Scattering (SAXS), which is 
sensitive  to both composition and density fluctuations. Silica glass is a
good candidate to study density fluctuations as composition fluctuations
do not occur in this ``single component'' glass (made of SiO$_2$ 
 ``molecules''). It represents a material of
significant commercial as well as fundamental interest. The characterization
of density fluctuations in silica is of great importance for the fundamental
understanding of the glass transition but also for commercial applications,
such as for silica-based optical fibers.

Density fluctuations are related to thermodynamics parameters such as
 compressibility of the sample. Small angle x-ray scattering intensity
measurements in silica glass across the glass transition temperature give
access to the isothermal compressibility. It can be calculated in the melt,
 as well as 
in the glassy state where it depends on the thermal history of the sample. 
In the glassy state, the isothermal compressibility can be decomposed
 into different terms  contributing to frozen-in or still propagating
 density fluctuations.

In this paper we will use  SAXS intensity measured both as a function of 
temperature and  fictive temperature, in combination with previous 
Brillouin scattering data,  \cite{LeParc2005}  in order to 
determine the different compressibility contributions to the density
fluctuations, both above and below the glass transition temperature. 
By a full point by point analysis, we obtain the temperature and fictive 
temperature variations of theses different compressibility contributions.
The results are discussed in comparison with previous  measurements.

\section{Theoretical background}

The theoretical description of the SAXS intensity for one composition unit or molecule,
 $I(q)$, due to density 
fluctuations
is rather well-known in liquids. Thermodynamic fluctuations theory predicts
that: 
\begin{equation}
I(q=0)=\frac{N_a \times (\sum Z)^2 \times \rho_0 }{m} \times k_b T \chi^0_T(T),
\label{eqn:comp}
\end{equation}
where $N_a$ is Avogadro's number, $\sum Z$ the total number of electrons per
molecule, $m$ the molar mass of one molecule, $\chi^0_T(T)$  the
isothermal compressibility of the liquid, $\rho_0$  the density and $k_b$
the Boltzman's constant. A similar  equation can  be found for the low $q$ limit of the static structure factor:

\begin{equation}
S(q=0)=n \times k_b T \chi^0_T(T)=\frac{N_a \times N \times \rho_0}{m} k_b T \chi^0_T(T)
\label{eqn:compsq}
\end{equation}
where $S(q)$ is the static structure factor,  $n$ is the number 
density of atoms and  $N$  the number of atoms  in the molecule.\cite{Buchenau}
Then $S(q=0)$ is  related to the intensity $I(q=0)$ per 
molecule by $S(q=0)=N   I(q=0)/(\sum Z)^2$.

In the glassy state, equation (\ref{eqn:comp}) is no more valid because
 the system 
is out of thermodynamic equilibrium. Historically, several different
 descriptions have been proposed to relate  the
low $q$ SAXS or light scattering
intensity  to compressibilities in the glassy state.

Weinberg first proposed that all density fluctuations are frozen-in 
at $T_g$, and
thus, equation (\ref{eqn:comp}) becomes independent of the temperature:\cite
{Weinberg1963}
\begin{equation}
I(q=0)=\frac{N_a \times (\sum Z)^2 \times \rho_0 }{m} \times k_b  T_g
\chi^0_T( {\bf T_g}).  \label{eqn:comp2}
\end{equation}

Wendorff rather said that below the glass transition,\cite{Wendorff1973}
\begin{equation}
I(q=0)=\frac{N_{a}\times (\sum Z)^{2}\times \rho _{0}}{m}\times k_b 
{\bf T}\chi_{T}^{0}(T_{g}),  \label{eqn:comp3}
\end{equation}
which yields to density fluctuations which are still decreasing with
decreasing temperature, but should go to zero at absolute zero temperature.

Roe and Curro mentioned that below the glass transition temperature, the
density fluctuations amplitude is much higher than that expected from (\ref
{eqn:comp3}). They assumed that the density fluctuations can be decomposed
into two components: one dynamic contribution, related to the isothermal
compressibility in the glass, which is determined by equation (\ref
{eqn:comp3})\cite{Roe1983,Curro1984} and one static or quasi-static 
term, $I(q=0,T=0)$, which represents the density
fluctuations that are frozen-in. This static contribution can be
attributed to structural entities whose relaxation times are longer than
experimental time scale. 
\begin{equation}
I(q=0)=I(q=0,T=0)+\frac{N_a \times (\sum Z)^2 \times \rho_0 }{m} 
 k_b{\bf T} \chi_T^0(T_g)
\label{eqn:comp4}
\end{equation}
\hskip 5.5cm static \hskip 2.5cm dynamic

 On another hand, 
Ruland and coworkers decomposed the density fluctuation below $T_{g}$ in one
component due to frozen-in disorder, independent of temperature, 
$I(q=0,T=0)$, and one temperature dependent component due to pressure fluctuations
 (longitudinal elastic waves),
\begin{equation}
I(q=0)=I(q=0,T=0)+\frac{N_{a}\times (\sum Z)^{2}\times \rho _{0}}{m}
k_bT\langle \frac{1}{\rho V_{l,\infty}^{2}}\rangle \label{eqn:comp5}
\end{equation}
where $V_{l,\infty}$ is the longitudinal sound wave velocity and $\langle 
 \rangle$ indicates a spatial average.\cite
{Wiegand1979,Rathje1976}

The same kind of decomposition has been introduced by Laberge using in equation (1) the
adiabatic compressibility, 
$\chi _{S}^{0}$:  $\chi_T^0=(\chi_T^0-\chi_S^0)+\chi_S^0$. 
The fluctuations can be
separated in isobaric fluctuations, related to $(\chi _{T}^{0}-\chi
_{S}^{0})$ 
and adiabatic fluctuations, related to $\chi _{S}^{0}$.\cite{Laberge1973}

The adiabatic fluctuations consist of sound waves and variations in the
local structure.  Laberge\cite{Laberge1973} introduced a term of 
relaxational origin in the
density fluctuations by pointing out that if the material is viscous enough
so that the structural relaxation is slow compared to the oscillations of
infinite frequency acoustic waves, then, the phonons contribution is defined 
by the infinite
frequency compressibility $\chi _{S}^{\infty }$ and the residual part by the
relaxational compressibility $\chi _{S}^{r}=\chi _{S}^{0}-\chi _{S}^{\infty
} $. 
These assumptions  lead to the following equation:

\begin{equation}
I(q=0)=\frac{N_a \times (\sum Z)^2 \times \rho_0 }{m} k_bT \left[
 (\chi_T^0-\chi_S^0)+
\chi_S^{r}+\chi_{S}^{\infty} \right], 
\label{eqn:comp6}
\end{equation}

with:
 \begin{equation}
\chi_S^{\infty }=1/(\rho (V_{l,\infty})^2-4\rho (V_{t,\infty})^2/3)=
1/(M_{\infty}-4G_{\infty}/3)=1/K_{\infty}
\label{eqn:chinf},
\end{equation}

where $V_{t,\infty}$ is the high frequency  transverse sound velocity, 
$G_{\infty}=\rho 
(V_{t,\infty})^{2}$ is the high frequency shear modulus,
 $M_{\infty}=\rho
(V_{l,\infty})^{2}$, the high frequency longitudinal modulus and 
$K_{\infty}$ is the infinite frequency compression modulus.

He, moreover, replaced $\chi_S^{\infty}$ by   $M_{\infty}^{-1}$, and 
$\chi _{S}^{r}$ by $\chi _{S}^{0}-M_{\infty }^{-1}$, considering 
that this is more appropriate in a viscoelastic material capable 
of supporting high-frequency shear stress.
The scattered amplitude due to density  fluctuations then becomes:

\begin{equation}
\frac{I(q=0)}{N_{v}\times (\sum Z)^{2}}=k_bT\left[ (\chi
_{T}^{0}(T)-\chi _{S}^{0}(T))+(\chi _{S}^{0}(T)-M_{\infty }^{-1}(T))+
M_{\infty }^{-1}(T)\right]\label{eqn:Labliq},
\end{equation}
with $N_v=N_a \times \rho_0 /m$.
This equation is valid for a viscoelastic medium in equilibrium
 at temperature $T$.

In the glassy state, Laberge considered that the  two first terms, describing 
non-propagating fluctuations, are kinetically arrested at a temperature in 
the annealing range because their
 relaxation time becomes considerably longer than the experimental time,  
as temperature is lowered. He assumed that these frozen-in fluctuations 
are the equilibrium fluctuations
 present in the 
glass at the fictive temperature $T_f$, thus giving:

\begin{equation}
\frac{I(q=0)}{N_{v}(\sum Z)^{2}}=k_bT_{f}[\chi
_{T}^{0}(T_{f})-\chi_{S}^{0}(T_{f})]+k_bT_{f}[\chi _{S}^{0}(T_{f})-M_{\infty }^{-1}(T_f)]+k_bTM_{\infty }^{-1}(T).
\label{eqn:Laberge}
\end{equation}
\hskip 4.5cm term 1 \hskip 3.5cm term2 \hskip 2.7cm term 3

 The third term is of vibrational origin and is identical to  
 the phonon term in equation (\ref{eqn:comp5}). The first two terms are frozen-in.
It can be noted that    $T_g$ or  $T_f$ are  used indifferently to 
characterize 
the temperature  at which a glass is frozen-in.  In the following, 
we use $T_f$ for samples with a known thermal history (annealed and 
equilibrated at $T_f$) and $T_g$ by
 reference 
to a given composition when the thermal history is not known. The 
temperature dependence
 below  the glass transition is related to term 3, proportional 
to temperature.

The fluctuations at the nanometer scale can also be probed by light
scattering. Using Brillouin light scattering, the first two terms 
 occurs in the elastic part \textit{i. e.} the Rayleigh part,
whereas the third one, related to the longitudinal velocity of sound arises
from propagating fluctuations and then contributes to the inelastic (Brillouin)
intensity. Thus, the ratio of the static to propagating intensities, also
called the Landau Placzek ratio, $R_{LP}$, gives access to the ratio of the
first two terms over the third. In a viscoelastic liquid,\cite{BucaroDardy1974}  
using equation (\ref{eqn:comp}):
\begin{equation}
R_{LP}(T)=\frac{\chi _{T}^{0}(T)-M_{\infty}^{-1}(T)}{M_{\infty }^{-1}(T)}
=\chi _{T}^{0}(T)M_{\infty }(T)-1.
\label{eqn:RLP1}
\end{equation}

 In a glass,\cite{KrolLyons1986}, equation (\ref{eqn:Laberge}) yield:

\begin{equation}
R_{LP}(T)=M_{\infty }^{{}}(T)\left[ \chi _{T}^{0}(T_{f})-M_{\infty
}^{-1}(T_{f})\right] \frac{T_{f}}{T}.\label{eqn:RLP2}
\end{equation}

\section{Previous measurements}

There are relatively few results about the compressibility in silica 
and especially few results concerning 
its temperature  and fictive temperature dependence. Most
of them use light scattering. The different compressibility values quoted 
in the literature 
are reported in table \ref{tab:tab1}.

\begin{table}
\begin{tabular}{lllll}
\hline \hline
authors & $T$ or $T_f$& $\chi _{T}^0(T \mbox{ or})(\mbox{m s}^{2}\mbox{ kg}^{-1})$ 
& $\chi _{S}^{r}(\mbox{m s}^{2}\mbox{ kg}^{-1})$ &
 $\chi_{S}^{\infty }(\mbox{m s}^{2}\mbox{ kg}^{-1})$ \\ \hline 
Laberge et al\cite{Laberge1973}& $T_f=1600\mbox{  K}$ &
$7.3\pm 0.6\times 10^{-11}$ &  &  \\ 
	Bucaro et al\cite{BucaroDardy1974}&$T=1400-1700$ K& 
$8.5\pm 0.3 \times 10^{-11}$ & $
6.3\times 10^{-11}$ & $2.2\times 10^{-11}$ \\ 
Krol et al\cite{KrolLyons1986}& $T_f=1273\mbox{  K}$&$7.72\times 10^{-11}$ & 
$5.68\times 10^{-11}$ & $2.04\times 10^{-11}$ \\ 
Watanabe et al\cite{WatanabeSaito2003}& $T_f=1230-1670\mbox{  K}$&$10.5\times 10^{-11}$ & 
$6.4\times 10^{-11}$ & $4.1\times 10^{-11}$ \\ 
Polian et al \cite{Polian2002}&$T=1370\mbox{  K}$&&&$2.23 \times 10^{-11}$\\
this work and\cite{LeParc2005} & $T_f=1373\mbox{  K}$&$6.16\pm 0.24 \times 10^{-11}$ & 
$3.88 \pm0.25 \times 10^{-11}$ & $2.28\pm0.07\times 10^{-11}$\\
this work and\cite{LeParc2005} & $T_f=1773\mbox{  K}$&$5.69\pm 0.23 \times 10^{-11}$ & 
$3.46 \pm0.22 \times 10^{-11}$ & $2.22\pm 0.07\times 10^{-11}$ \\\hline \hline
\end{tabular}

	\caption{\label{tab:tab1}Comparison of isothermal compressibility, adiabatic relaxational compressibility and adiabatic infinite frequency compressibility taken from the literature. }

\end{table}

Bucaro et Dardy used Brillouin light scattering and measurements of the
Landau Placzek ratio to determine the isothermal compressibility 
$\chi _{T}^{0}$ at high temperature,
between 1400 and 1700K, using equation (\ref{eqn:RLP1}).\cite{BucaroDardy1974}
 This
equation is only valid  in the (viscous) liquid. They found $\chi
_{T}^{0}=8.5\pm0.3\times 10^{-11}\mbox{ m s}^2\mbox{  kg}^{-1}$with no temperature
dependence in the range where $\chi_{T}^{0}$ was determined. 
They also deduced the
compressibility at infinite frequency $\chi _{S}^{\infty }$ 
directly from the longitudinal and transverse Brillouin shifts; they obtained 
 about 
$2.2 \times 10^{-11}\mbox{ m s}^2\mbox{  kg}^{-1}$ in the same temperature range.

Krol et al\cite{KrolLyons1986} also measured the Landau Placzek ratio as
 well as
Brillouin shifts (shear and longitudinal modulus) as a function of
temperature in a silica glass with a fictive temperature equal to
\mbox{1273  K}. They used equation (\ref{eqn:RLP2}), taking
 into account
the variation of the high frequency modulus with temperature. From
the slope of $R_{LP}(T)/M_{\infty }(T)$, plotted below the glass transition
 as a function of $1/T$, they deduced 
$\chi_{T}^{0}(T_{f})$. Calculating the infinite frequency adiabatic 
compressibility from both
longitudinal and transverse Brillouin shifts, 
$\chi_S^{\infty }=(M_{\infty}-\frac{4}{3}G_{\infty })^{-1}$, they deduced the 
relaxational contribution at 
$T_{f}$, $\chi_{S}^{r}(T_{f})=\chi_{S}^{0}-\chi_S^{\infty}\simeq 
\chi_{T}^{0}-\chi_S^{\infty}$
 because in
silica $\chi_{S}^{0}-\chi_{T}^{0}$ is negligible. Indeed, it is
 related to thermal expansion,\cite{Laberge1973} which is very low in silica glass:
 $\chi_T^0(T_f)-\chi_S^0(T_f)=\frac{\alpha(T_f)^2 \times T_f}
{\rho_0 \times C_p(T_f)}$,
$T_f
=1273\mbox{ K}$, $\alpha(T_f)=0.5 \times 10^{-6} \mbox{ K}^{-1}$ 
(Ref. ~\onlinecite{Bruckner1970}),
  $\rho_0=2202\mbox{ kg m}^{-3}$, 
and $C_p(T_f)=
1231\mbox{ m}^2\mbox{ s}^{-2}\mbox{ K}^{-1}$. These values yield to 
$\chi_T^0(T_f)-\chi_S^0(T_f)=1.08 \times 10^{-16} \mbox{  m s}^2\mbox{ kg}^{-1}$
 Then,  Krol et al obtained 
$\chi_{T}^{0}(T_f)=7.72\times 10^{-11}\mbox{ m s}^2\mbox{  kg}^{-1}$,  and $\chi
_{S}^{\infty }=2.04\times 10^{-11}\mbox{ m s}^2\mbox{  kg}^{-1}$, from which they 
deduce $\chi_{S}^{r}
\simeq 5.68\times 10^{-11}\mbox{ m s}^2\mbox{  kg}^{-1}$.

Saito et al\cite{SaitoKakiuchida1997,SaitoIkushima1997} deduced the 
isothermal compressibility from
the light  scattering intensity, $I_{\mbox {\small light}}$, as a function 
of temperature.
They used equation (\ref{eqn:comp}) in the liquid and the following 
expression in the glass: 
\begin{equation}
I_{\mbox {\small light}}\simeq
I(q=0)\propto \chi _{T}^{r}(T_{f})kT_{f}+
\chi_{S}^{\infty }(T)kT,\label{eqn:Saito}
\end{equation}
where $\chi_{T}^{r}=\chi _{T}^{0}-\chi_{T}^{\infty}$. This equation differs 
from equation (\ref{eqn:Laberge}) by the use of $\chi_{S}^{\infty }$
instead of $M_{\infty}^{-1}$ as the temperature-dependent part,  the shear term
 being small enough to be neglected according to the authors.
 Another difference with equation (\ref{eqn:Laberge}) is the use of
$\chi_{T}^{0}-\chi_{T}^{\infty}$ 
instead of $\chi _{T}^{0}-\chi_{S}^{\infty}$.
  They  found a
sudden increase of the isothermal compressibility at the glass transition
and at the melting temperature, with no measurable temperature dependence within the
glassy and supercooled liquid state.\cite{SaitoKakiuchida1997} They found 
$\chi_T^{0 }$ around $1.8 \times 10^{-11}\mbox{ m s}^2\mbox{  kg}^{-1}$ 
in the glass,  and around $8 \times 10^{-11}\mbox{ m s}^2\mbox{  kg}^{-1}$in 
the supercooled liquid, which yields  to $\chi_T^r$ 
around $6 \times 10^{-11}\mbox{ m s}^2\mbox{  kg}^{-1}$.\cite{SaitoIkushima1997}

Apart from Saito's measurements, most  compressibility measurements 
 as a function of temperature across the glass transition concern the 
 infinite frequency compressibility.   $\chi
_{S}^{\infty }$  has been measured  using Brillouin 
light scattering by Polian et al\cite{Polian2002} as a function of
 temperature using Brillouin scattering
measurements of the longitudinal and transverse modes. They observed 
that $\chi _{S}^{\infty }$
decreases continuously  as a function of temperature from about 
$2.7 \times 10^{-11}\mbox{ m s}^2\mbox{  kg}^{-1}$ at $400 \mbox{ K}$
 to $2.05 \times 10^{-11}\mbox{ m s}^2\mbox{  kg}^{-1}$ at $2300 \mbox{ K}$. 
Those values 
are in agreement with high frequency previous data from several 
sources\cite{McSkimin,Fine1954,Spinner1956} 
compiled by Vukcevitch.\cite{Vukcevitch1972}  The data of Polian are also
 consistent with 
our determination of $\chi_{S}^{\infty }$ as a function of 
temperature.\cite{LeParc2005}

Concerning the effect of fictive temperature, Champagnon et al\cite
{ChampagnonLeParc2002} measured the Landau Placzek ratio at room temperature
in several silica samples of same origin but prepared with different fictive
temperatures from $T_{f}=1373 \mbox{ K}$ to $T_{f}=1773 \mbox{ K}$,  using heat-treatment
 in the glass transition region
(transformation range). 
They observed an increase of the $R_{LP}$ at room temperature as a function 
of fictive temperature.
 The increase is approximately linear above $
T_{f}=1473\mbox{ K}$. From relation (\ref{eqn:RLP2}), one can deduce 
that $\chi_{T}^0(T_f)$ and $M_{\infty}^{-1}(T_f)$
 vary slowly with $T_f$ between 1473 and $1773\mbox{ K}$.

There is  one paper where  compressibility is determined using SAXS intensity
measurements at room temperature in four silica samples with different
 fictive temperatures.\cite{WatanabeSaito2003} 
The authors obtained a linear variation of the  SAXS intensity  as a function 
of fictive temperature. 
Following equation  (\ref{eqn:Saito}),  
they  use  a linear regression and obtained $\chi
_{S}^{\infty }$  assumed to be independent of temperature, from the slope, and 
$\chi _{T}^{r}$  at the fictive temperature, from the intercept. 
 They
obtained $\chi
_{S}^{\infty }=4.1 \times 10^{-11}\mbox{ m s}^2\mbox{  kg}^{-1}$ and 
$\chi_{T}^{r}(T_f)=6.4 \times 10^{-11}\mbox{ m s}^2\mbox{  kg}^{-1}$
for $\chi _{T}^{r}$. They add both terms
to  determine the adiabatic compressibility 
$\chi_T^0\simeq \chi _{T}^{r} +\chi
_{S}^{\infty }= 10.5 \times 10^{-11}$m 
kg$^{-1}$ s$^{2}$.

\section{Measurements}

We analyzed high purity silica samples (with  very low OH 
 content), all from the same batch. They are fusion glasses
of type I.\cite{Bruckner1970} The different samples have 
different well-defined thermal histories. 
The temperature, time  and conditions of annealing  are reported on table \ref{tabsamples}, 
along with the quenching conditions. 
Annealing times are estimated from the expected relaxation times at 
the temperature of the heat-treatment, 
and the quenching rate (specially for the high temperature treatments) are assumed to 
be fast enough to preserve the high temperature structure.
We plotted several spectroscopic features based on infrared or Raman 
measurements (position or intensities of peaks) as a function of annealing 
temperature. \cite{Champagnon2000,LeParcChampagnon2001}
Those spectroscopic features have been previously reported to be  attributed to structural 
characteristics sensitive to the fictive temperature, 
and to vary linearly with the inverse 
fictive temperature. \cite{Agarwal1955,Galeener1983} A linear relation
 was found  for  the six annealed samples,
using the temperature of treatment as fictive temperature.  Thus, we assumed that 
  all the  samples are treated 
long enough and quenched fast enough 
to be stabilized  at their annealing temperature. In other words, the fictive temperature
 is assumed to be 
equal to the annealing  temperature.  Moreover, the sound velocity, measured using Brillouin 
scattering vary also linearly, as plotted versus the fictive temperature, 
in agreement with the
 hypothesis that samples
are stabilized.

\begin{table}
\begin{tabular}{ccccc}
\hline \hline
&Annealing&&&Quench \\
temperature& time&conditions && conditions \\ \hline
1373 K&60 h&in air&& in air on a metallic plate\\\
1473 K&1h&in air&&in air on a metallic plate\\
1573 K&1h45&in dry air&&quenched in water\\
1773 K&1h45&in dry air&&quenched in water\\ \hline\hline 
\end{tabular}

\caption{Experimental conditions of annealing and quenching of the  silica samples.}\label{tabsamples}
\end {table}
The small-angle x-ray measurements were performed on the D2AM experimental
set-up of the European Synchrotron Radiation Facility (ESRF) at Grenoble,
France. The measurements were performed on small plates of about 1 mm
thickness. The data on silica samples were measured with an incident energy
of 18 keV, yielding to an available $q$-range
 of $0.02-1\mbox{ \AA}^{-1}$, where $q = 4 \pi \lambda^{-1} \sin \theta$
 is the scattering vector,
and $\theta$
is half of the scattering angle.
The accumulation times were of 200 s, and the data were collected using a
CCD camera. In situ temperature
measurements were performed using a high temperature molybdenum furnace
already referred to in  Ref.\ ~\onlinecite{Soldo1998}. The use of this furnace
 produces a
reduction of the available $q$-range,  which is about $0.02-0.7\mbox{ \AA}^{-1}$. 
The temperatures in the furnace were calibrated by 
observing the melting of a gold foil, which deviates from less than 2K from
 the standard value. Radial integration and corrections from  cosmic 
rays have been
performed in the usual way. The x-ray was monitored before the furnace, 
$m_0$, 
and after the furnace, $m_1$,
 by scattering part of the beam
with Kapton foils onto scintillation counters. Before measuring a new 
sample, the SAXS signal of the empty
furnace, $I^B(q)$, was systematically measured.    This background 
originated mainly from the Kapton 
windows and air scattering.

Based on the measured signal with the sample in place,
$I^{S+B}(q)$, the signal originating from the sample was calculated 
according to
\begin{equation}
I(q) \propto  (t \ln 1/t)^{-1} [I^{S+B}(q)/m_1^{S+B} - I^{B}(q)/m_1^{B}] 
m_1^{S+B} / m_0^{S+B},
\end{equation}
where the transmission factor $t = m_1^{S+B} m_0^{B} / (m_0^{S+B}  m_1^{B})$
allows for changes in the thickness and orientation of the sample. Doing 
this, the signal
 is corrected from the variations of the
incident flux, as well as normalized to the absorption of
the sample. The absolute intensity of the samples is
determined by normalization to a reference sample of pure water of 1 cm
thick. The scattering power of the water sample is taken equal to
6.37 electron unit per molecule unit H$_{2}$O (this 
value calculated  at 300K from equation (1) is in very good agreement 
with the 
measured value of 6.4$\pm 0.2$ electron unit
per molecule  H$_{2}$O).\cite{LevelutGuinier1967} The error in the final 
absolute 
intensity is estimated to be about $\pm2$\%.

\begin{figure}
\epsfxsize=350pt{\epsffile{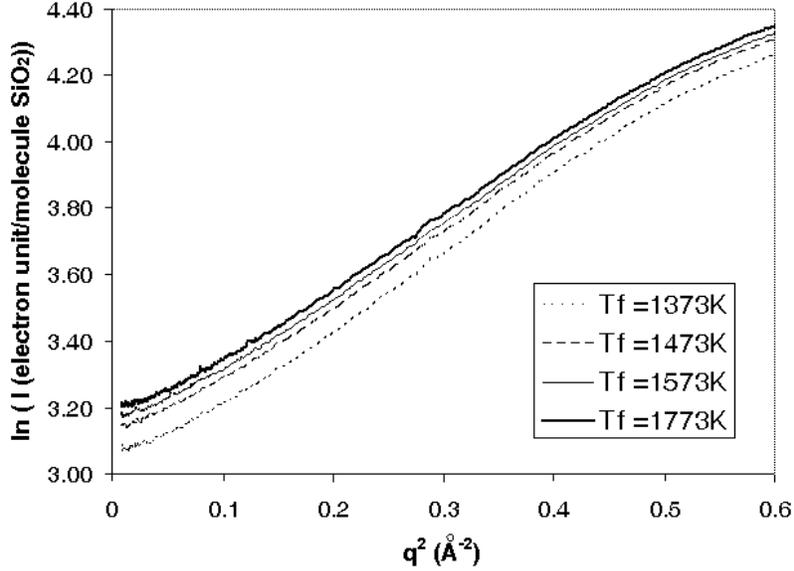}}
\caption{\label{fig:ibrute} Logarithm of the scattered intensity as a
 function of
 the square of the scattering vector, $q$, measured at room temperature
for several silica samples with different 
fictive temperatures, showing a linear regime  extending from
$q^2 \approx 0.04$ to 0.4  \AA$^{-2}$. This linear regime is used to 
extrapolate the
 intensity towards zero scattering vector, using $I(q) = I(q=0) \exp (bq^2)$. }
\end{figure}

\begin{figure}
\epsfxsize=250pt{\epsffile{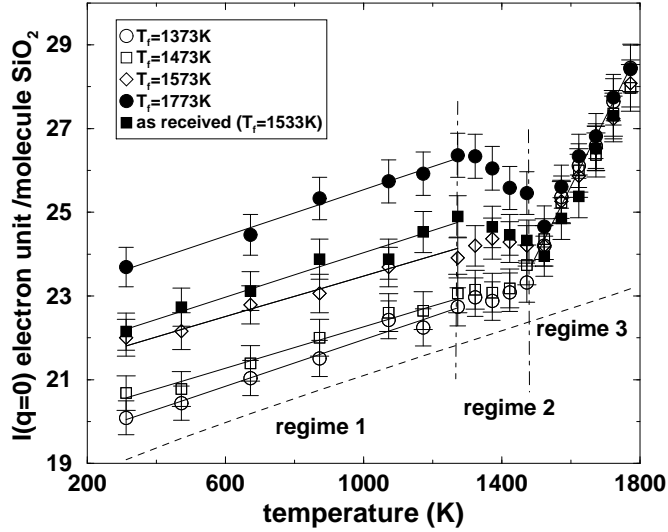}}

\caption{\label{fig:saxsintabs} Scattered intensity extrapolated to
 q=0 as a function of temperature, 
for five samples having different fictive temperatures. The temperature dependent term due to phonons, $I^{{\mbox {\small ph}}}$, calculated from  Brillouin measurements of Ref ~\onlinecite{LeParc2005},
is plotted  (dashed line) to show the slope of this contribution versus temperature. It is shifted of 18 e.u./molecule for clarity.}
\end{figure}

An example of scattering curves $I(q)$ versus $q$, measured at room
 temperature, 
 for four different samples, is shown on Fig.\ \ref{fig:ibrute}. 
The scattering curve, plotted as $\log(I(q))$ versus $q^2$ exhibits  a 
linear regime extending from
$q^2 \approx 0.04$ to 0.4  \AA$^{-2}$, which means that the intensity is well
described by $I(q) = I(q=0) \exp (bq^2)$ (Ref.\ ~\onlinecite{Wiegand1979})
 and $I(q=0)$ is obtained
as one of the regression parameters. The obtained values of $I(q=0)$ for a
 set of 5 samples with
different fictive temperatures are shown in  Fig.\ \ref{fig:saxsintabs}
 as a function of temperature.

 For one sample, three regimes can  generally be
identified.\cite{LevelutFaivre2002} The first one is an almost linear 
regime in the glassy state ($T<<T_g$), with 
a slight increase with increasing
temperature. The values of  intensity in this
regime depends on the fictive temperature of the sample. The third 
regime corresponds also to a quasi-linear 
increase but with a larger
slope in the supercooled liquid state ($T>T_g$). In this regime,
$I(q=0)$ values are independent of fictive 
temperature within experimental errors. In the intermediate
 temperature range  around the glass transition 
 $I(q=0)$ follows a  
``structural relaxation'' regime. The
shape of  curves  in this second regime depends on $T_{f}$ 
 as well as on the acquisition time.

\section{Compressibility}

In this section we present   compressibility results obtained with a
 full point by point analysis. 
From our set of SAXS measurements as a function of temperature and fictive
temperature we analyze the extrapolated SAXS intensity values, $I(q=0)$, 
in two regions. 
In the first  one corresponding to the glassy state regime (regime 1), 
we used   relation (\ref{eqn:Laberge}). In the 
second one corresponding to the supercooled
liquid (regime 3 as defined in previous paragraph), 
 relation (\ref{eqn:comp}) is used. Both regions 
corresponds to almost linear
regimes for the intensity $I(q=0)$ versus temperature. The compressibility 
cannot
be determined in the structural relaxation regime (regime 2), because the 
fictive
temperature in that region is not known. Indeed, in this regime, it 
is assumed that the fictive temperature  can  evolve
 during the time of the measurement.

\begin{figure}
\epsfxsize=250pt{\epsffile{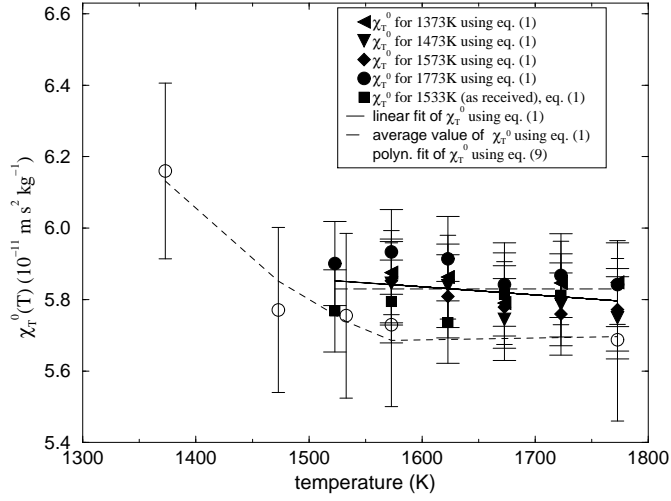}}

\caption{\label{fig:betaT}Isothermal compressibility  at equilibrium,
 determined at $T_f$ 
(filled symbols)  for five silica  samples having different 
thermal histories. $\chi_T^0$  has been determined 
using equation (\ref{eqn:comp}) in the liquid state and using 
equation (\ref{eqn:Laberge}), taking into 
account $M_{\infty}(T_f)$ deduced from Brillouin 
scattering measurements, in the glass state. The solid line in the 
supercooled liquid range, $1373-1773\mbox{  K}$, is a linear fit of the five sets
 of data above $1473\mbox{ K}$.
The dot-dashed line is a quadratic fit, shown as a guide for eyes, 
of $\chi_T^0(T_f)$ in the glass state. The long dashed  line is the average of the
values in the supercooled liquid, 
obtained from the point by point analysis. For clarity, the error bars have been plotted
 only for one sample in the supercooled liquid.}
\end{figure}

In the liquid state, above $T_g$, the isothermal intensity is simply 
calculated for
 each data point  by dividing the extrapolated SAXS intensity by the 
 temperature of the measurement:

\begin{equation}
 \chi_{T}^{0} =\frac{I(q=0)}{T \times N_{v}k_{b}(\sum Z)^{2}},
\end{equation}

using $\sum Z=30$, $\rho_0=2200 \mbox{ kg m}^{-3}$, 
$m=60 \times 10^{-3}\mbox{ kg mol}^{-1}$.
The error on the temperature in this range is less than $\pm 0.1 \%$, 
and consequently 
the error on the compressibility  is assumed to be equal to  the error 
on the intensity $I(q=0)$, that is about $\pm 2$\%. The values of $\chi_T^0$
 obtained from the $I(q=0)$ 
values for the different silica samples above $T_g$ are plotted as a function
 of temperature in fig. \ref{fig:betaT}. $\chi_T^0$ appears to decrease 
slightly with temperature
though the variations are within the errors bars. The mean value 
of $\chi_T^0$ calculated in this temperature range for the different silica samples
is  $5.82\pm 0.12 \times 10^{-11}\mbox{ m s}^2\mbox{  kg}^{-1}$.

In  order to determine the different contributions to 
density fluctuations in the glassy state, we use equation 
(\ref{eqn:Laberge}), where terms 1 and 2 have been 
combined:

\begin{eqnarray}
I(q=0) =&k_{b}N_v(\sum Z)^2 \left[ \chi_{T}^{0}(T_{f})-M_{\infty}^{-1}(T_f)
 \right] T_{f}&+ k_{b}N_v(\sum Z)^2 \left[  M_{\infty}^{-1}(T)  \right] T \nonumber \\  
=&I^{\mbox {{\small res}}}&+I^{\mbox {{\small ph}}}. 
\end{eqnarray}

We first focus on the temperature dependent term 
of equation (\ref{eqn:Laberge}), 
 $I^{\mbox {{\small ph}}}$.
 This contribution (term 3),
   related to the longitudinal phonons, can be calculated independently of 
our SAXS measurements from Brillouin light
 scattering experiments. In order to determine the longitudinal velocity at 
 the precise temperatures at which the SAXS measurements
have been performed,  we interpolate  our previous  high 
resolution Brillouin 
measurements,\cite{LeParc2005} performed from 300 to $1773\mbox{  K}$  using 
the following polynomial fit:
\begin{equation}
V_{l,\infty}\mbox{(m/s)}=5788 +0.6578\times T -0.0001844\times T^2 \pm 0.5.
\end{equation}

The sound velocity has been calculated from the Brillouin shift 
using a  temperature dependence of 
 refractive index. We used a linear 
variation of the index $n$ with temperature which does not differ 
from more than 0.2\% from the temperature variations $\Delta n$  
 measured by  Br\"uckner.\cite{Bruckner1970}
The temperature dependent index estimated by Polian et al\cite{Polian2002} 
is rather similar to that of Br\"uckner up to 973K but the index calculated 
by Polian is smaller above. The maximum difference is about $\pm 0.25$\% 
(and this would lead to less than  
about $\pm 0.5$\% difference on the determination of  the intensity 
$I^{\mbox  {\small ph}}$). In order to calculate $M_{\infty}^{-1}$, we used a 
temperature dependent density determined using either measurements of 
specific volume by Br\"uckner or experimental
 values of the linear thermal dilatation compiled by him.\cite{Bruckner1970}
 Both methods yield to density which vary from 
less than 0.2 \% in the considered temperature range and the differences on 
$M_{\infty}^{-1}$
and $I^{{\mbox {\small ph}}}$ are accordingly negligible. The error  on 
 $I^{{\mbox {\small ph}}}$ is about 0.5  \% (0.1\% on the 
temperature, 0.4\% on $M_{\infty }^{-1}$). Moreover,
 the variation of $V_{l,\infty}$ with fictive temperature is less than 
0.6 \% (1.2\% on 
$M_{\infty }^{-1}$).   

\begin{figure}
\centerline{
\epsfxsize=210pt{\epsffile{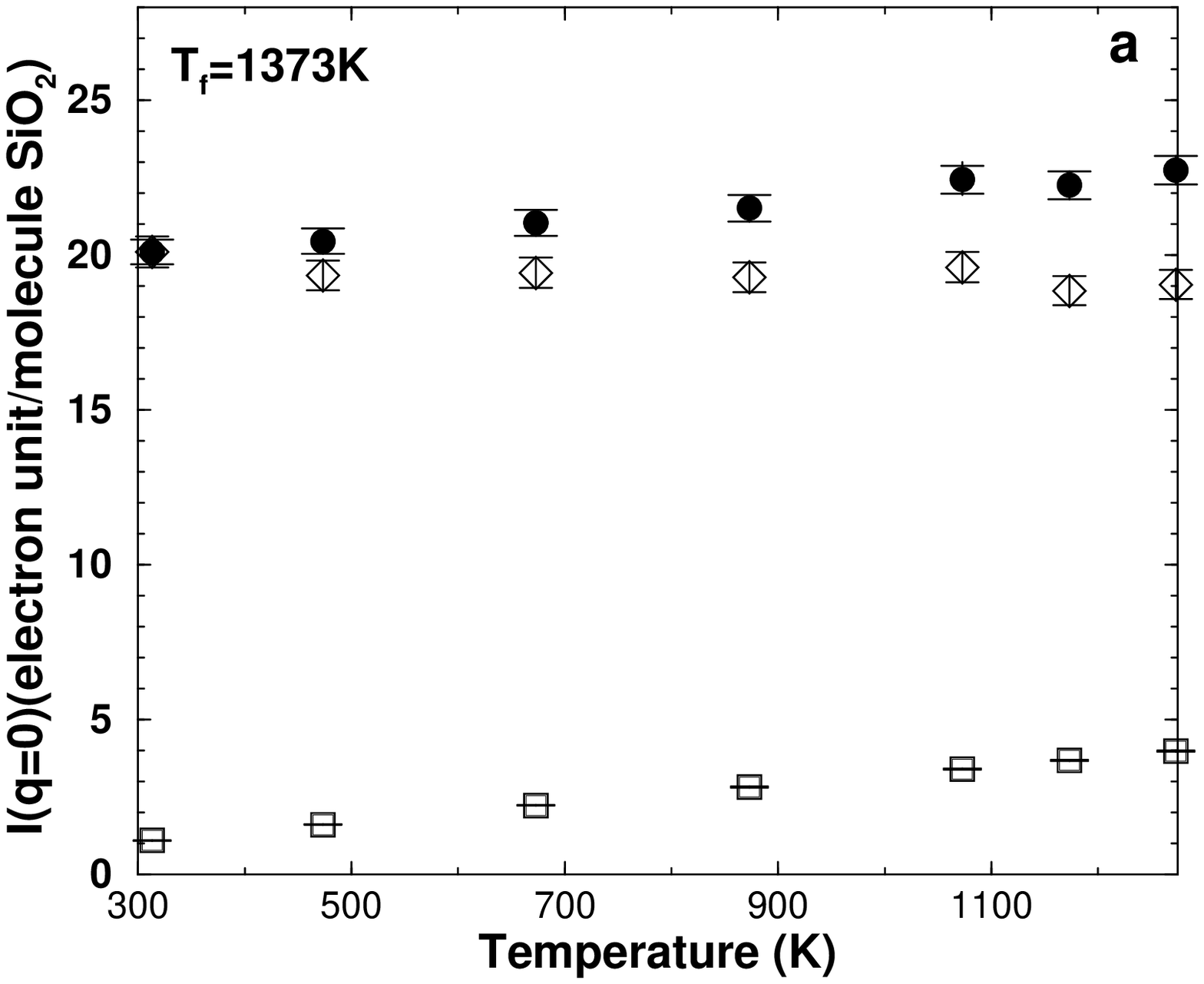}}\hspace{0.5cm}
\epsfxsize=210pt{\epsffile{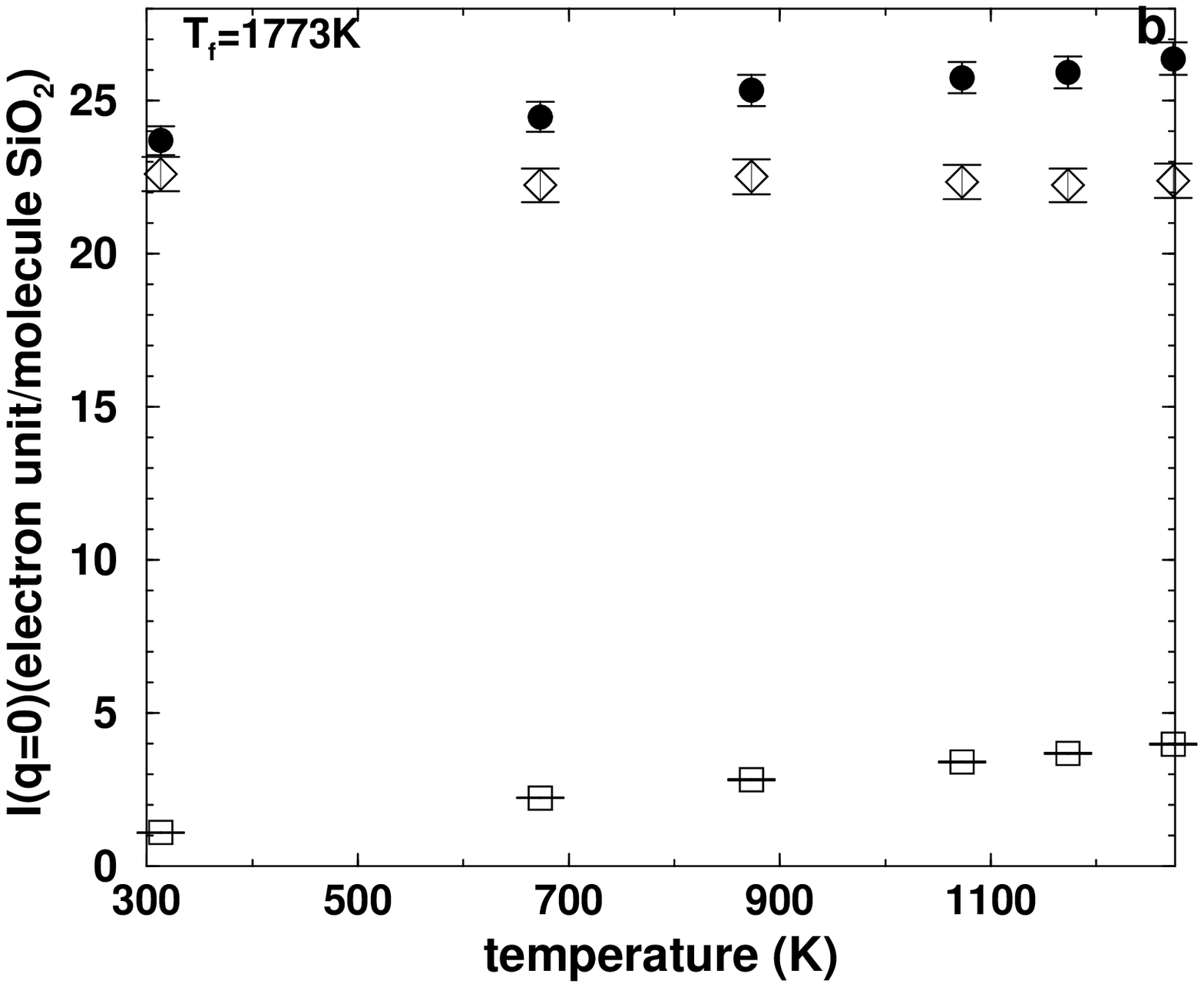}}} 
\caption{\label{fig:contr}(a) Scattered intensity extrapolated to $q=0$
 ($\bullet$), $I(q=0)$, in the supercooled liquid 
state for the sample of lowest fictive temperature, $T_f=1373$K. The 
two contributions of equation (\ref{eqn:Laberge}), 
$I^{{\mbox {\small ph}}}$ ($\square$) due to phonons (term 3) and 
$I^{{\mbox {\small res}}}$ 
($\circ$)(related to terms 1 and 2) are also plotted. (b) Same 
 quantities for the sample with the higher fictive temperature, 
$T_f=1773\mbox{ K}$. }
\end{figure}

$I^{{\mbox {\small ph}}}$ increases with temperature, 
from  1 electron 
unit/molecule at room temperature to 4 electron 
units/molecule at $1300\mbox{ K}$, as observed in Fig.\ \ref{fig:contr}.
The contribution
 $I^{{\mbox {\small ph}}}$ is rather small, 5-20\% of $I(q=0)$. This result
 correlate rather well with that of Laberge which mentioned that this term  
contribute to about 8\% of the total scattered intensity.\cite{Laberge1973}
However, the increase of this term should be  responsible  for the increase of $I(q=0)$ 
with increasing temperature
 in the glassy state, as stated by Laberge. The variation of
 $I^{{\mbox {\small ph}}}$ with 
fictive temperature has been neglected because it amounts to less than 
0.25\% of the total intensity because of the small contribution of the
 phonons term to $I(q=0)$.

Subtracting $I^{{\mbox {\small ph}}}$ from the 
 total extrapolated SAXS intensity in the glassy state  gives access to 
the residual contribution, proportional to $ \left[ 
\chi_{T}^0(T_f)-(M_{\infty}(T_f))^{-1} \right]  T_{f}$. 
For a given fictive temperature, $I^{{\mbox {\small res}}}$
very slightly decreases with  temperature, by about  2 to 5\% from 
300 to 1300K, depending on the sample. This decrease is systematic for the 5 different fictive temperature
 silica samples that we analyzed, but at the limit of accuracy of the experiment. However, 
we note that other methods of analysis (for example using $\chi_S^{\infty}$ for the temperature
 dependent term) would yield to a stronger decrease.
 $I(q=0)$, $I^{{\mbox {\small ph}}}$ 
and $I^{{\mbox {\small res}}}$
are presented on Fig. \ref{fig:contr} for  $T_f= 1373\mbox{ K}$ (Fig. 
\ref{fig:contr}(a)) and for $T_f= 1773\mbox{ K}$ (Fig. \ref{fig:contr}(b)).

The residual contribution, $I^{{\mbox {\small res}}}$, yields 
$\chi_{T}^{0}(T_{f})-M_{\infty}^{-1}(T_f)$.
 The accuracy on the residual contribution is $\pm2.5$\%,
 but the error in fictive temperature ($\pm 10\mbox{ K}$)  
 has to be taken into account 
to estimate the error on $\chi_{T}^{0}(T_{f})-M_{\infty}^{-1}(T_{f})$ which 
is then $\pm 3.5$\%. 
Then  $\chi_{T}^{0}(T_{f})$ can be calculated in the glassy state using
 $M_{\infty}^{-1}(T_{f})$, 
deduced from Brillouin measurements.  $\chi_{T}^{0}(T_{f})$, also presented in 
Fig.\ \ref{fig:betaT}, varies  
  between  $ 6.16\times 10^{-11}  \mbox{ m s}^2\mbox{  kg}^{-1}$ for 
$T_f=1373\mbox{ K}$ to   $5.69\times 10^{-11}  
\mbox{ m s}^2\mbox{  kg}^{-1}$  for $T_f=1773\mbox{ K}$, 
decreasing by 8\% in total with 
 increasing  fictive temperature. The error bar is  $\pm$ 4\% 
($\pm 0.24 \times 10^{-11}  \mbox{ m s}^2\mbox{  kg}^{-1}$). 
The values of $\chi_T^0(T_f)$
obtained in the glass state are 
in rather good agreement with
 the values obtained for the supercooled liquid. All the values
 are compatible with a smooth decrease of isothermal compressibility
with temperature.

 In order to understand the different contributions to 
the isothermal compressibility in the glassy state, as well as for
 comparison with data of the literature, it is interesting to calculate 
 the infinite frequency  and  relaxational compressibilities. 
The infinite frequency  compressibility can be determined from our previous
 Brillouin scattering measurements\cite{LeParc2005} using equation 
(\ref{eqn:chinf}).
 For this, we need a precise determination of  
 the high frequency transversal velocity. We used  the same temperature 
dependence for the refractive index as for the longitudinal velocity
 and found that our data in the $300-1773\mbox{  K}$ 
range can be well described by:

\begin{equation}
V_{t,\infty}\mbox{(m/s)}=3689 +0.2946 \times T -0.0001014 \times T^2 \pm 30
\end{equation}

\begin{figure}[h]
\epsfxsize=240pt{\epsffile{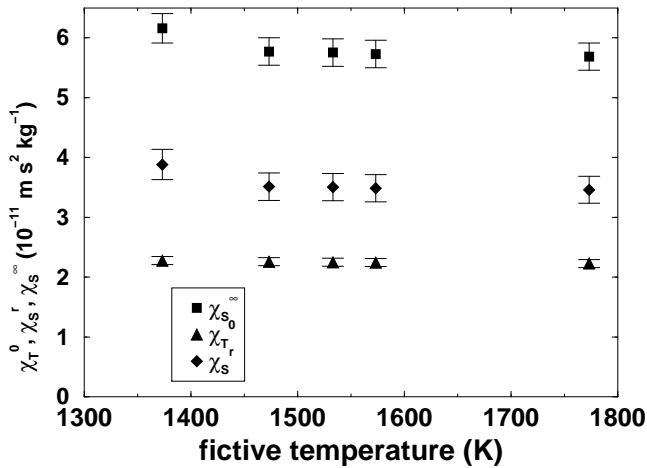}}

\caption{\label{fig:chiinfrelax} Isothermal compressibility in the
 glass state $\chi_T^0(T_f)$ as a function of fictive
 temperature, together with the infinite frequency compressibility 
$\chi_T^{\infty}(T_f)$ and the 
relaxational
 compressibility  $\chi_T^r(T_f)$.}

\end{figure}

The infinite frequency compressibility, is calculated from  the
 combination of the experimental
 determinations of longitudinal and transverse velocity using  
equation (\ref{eqn:chinf}).
  The error 
bar on the infinite frequency compressibility is estimated to be about 
3\% (0.2\% on $\rho$, 1\%  on the square of
 longitudinal sound velocity, 1.8\% on the square of transverse velocity) 
 {\mbox {\it i. e.}} $0.07 
 \times 10^{-11} \mbox{ m s}^2\mbox{  kg}^{-1}$. 
The error is larger on transverse velocity than on longitudinal one because
 it has to be measured using right angle geometry instead of backscattering 
used for longitudinal velocity.   We have neglected the very low variations 
of  the transverse and the longitudinal velocities with fictive temperature. 
The infinite frequency compressibility can also be fitted by a quadratic 
expression in $T$, valid in the $300-1773\mbox{  K}$ range:
\begin{equation}
\chi_S^{\infty}(\mbox{m s}^2\mbox{ kg}^{-1})=2.917\times 10^{-11}
 -7.309\times 10^{-15} \times T +1.925\times 10^{-18}
 \times T^2 \pm 0.07 \times 10^{-11}. \label{eqn:vtrans}
\end{equation}

At the high frequency of Brillouin scattering measurements, 
 the samples are in equilibrium and  we calculate $\chi_S^{\infty}(T_f)$ using the  values 
of $\chi_S^{\infty}(T)$ given by equation (\ref{eqn:vtrans}) at $T=T_f$. Those values are  plotted in Fig.\ 
\ref{fig:chiinfrelax} as a function of $T_f$.  $\chi_S^{\infty}$ 
 slightly decreases  with fictive  in this  range  we analyzed here: 
from $2.28  \pm 0.07 \times 10^{-11} \mbox{ m s}^2\mbox{  kg}^{-1}$ at $1373\mbox{  K}$ 
to $2.23  \pm 0.07 \times 10^{-11} \mbox{ m s}^2\mbox{  kg}^{-1}$ at $1773\mbox{  K}$, 
corresponding to a 2\% decrease.
 $\chi_S^{r}(T_f)$ is then obtained by subtracting
$\chi_S^{\infty}(T_f)$ to $\chi_T^{0}(T_f) \simeq \chi_S^{0}(T_f)$
 for the five samples with different 
fictive temperatures. The error on $\chi_S^{r}(T_f)$ is estimated to be within 
  6.5\%. The results are shown in Fig.\ \ref{fig:chiinfrelax}. The 
relaxational contribution 
varies  from $3.88 \pm 0.25 \times 10^{-11} \mbox{ m s}^2\mbox{  kg}^{-1}$ for $T_f=1373$K
 to $3.46 \pm 0.23 \times 10^{-11} \mbox{ m s}^2\mbox{  kg}^{-1}$ 
for $T_f=1500^{\circ}$C.  It is then slightly decreasing with fictive 
temperature, by about 10\% in the 1373--1773 range. In Fig.\ 
\ref{fig:chiinfrelax}, one can see that the infinite frequency 
compressibility contributes to  about 40\% of the static isothermal compressibility
 whereas  the dominant contribution is the relaxational one, which amount to around 60\%.
Our value of the relaxational compressibility is somewhat smaller 
than the other determinations of table \ref{tab:tab1}. However, 
the higher of those determinations from the literature,\cite{WatanabeSaito2003}
 identified as $\chi_{T}^{r}(T_{f})$ by the authors,
is rather  equal to $\chi_{T}^{0}(T_{f})-M_{\infty}^{-1}(T_f)$ 
 if we follow equation (\ref{eqn:Laberge}). Thus we  compared  their value
 directly to our value of 
$\chi_{T}^{0}(T_{f})-M_{\infty}^{-1}(T_f)$ which is equal to 
$5.03 \times 10^{-11} \mbox{ m s}^2\mbox{  kg}^{-1}$ 
for the sample of $T_f=1373\mbox{ K}$, decreasing down to   to 
$4.57 \times 10^{-11} \mbox{ m s}^2\mbox{  kg}^{-1}$ 
 for the others samples. The discrepancy is then smaller.

\section{Discussion}

The first result of this paper is a new determination of the isothermal 
compressibility, obtained using small angle X-ray scattering.
Most of the measurements of the isothermal compressibility quoted in the 
literature
 (see table \ref{tab:tab1})
have been established using light scattering.  It can be seen that the 
results for isothermal compressibility are rather
 dispersed. Our results are  in agreement with previous one within error
 bars, though 
slightly lower than  most of them.
 Part of the differences  can  occur from 
the experimental uncertainty on intensity measurements using light
 scattering technique. Indeed, 
 any defect in the sample, such as
unperfect surface polishing, bubbles and so on, can contribute in an
extrinsic way to the elastic scattering.  Measurements of intensity 
using  light scattering are even more difficult at high temperature 
(due to thermal radiation
 which increases the noise). Another difficulty with  light scattering 
 measurements is that it requires large 
samples, in order to minimize the parasitic contribution from the surface. 
  And on large samples, the homogeneity of 
the fictive temperature   is difficult to achieve. 
Bucaro et Dardy\cite{BucaroDardy1974} estimated an 
error of about $\pm 3\%$ on their measurements of 
 $R_{LP}$ ratio. Krol et al\cite{KrolLyons1986} 
 mentioned that several determinations of the Landau Placzek ratio differ 
from less than  2\%. 
From our own measurements, we estimate  that the Landau-Placzek ratio 
is usually measured with an  
accuracy of at least 
  $\pm 1$ {\it i.e.}  $\pm 4\%$ for silica at room temperature.
To illustrate this
 we calculated,  using equation  (\ref{eqn:RLP2}),  the isothermal 
compressibility  
starting from Landau-Placzek
 ratio measured with a matching index liquid.\cite{Champagnonprivate}
 We obtain 
$\chi_T^0= 9.8 \pm 3 \times 10^{-11}\mbox{m s}^2\mbox{ kg}^{-1}$.  The error bar
 contains dispersion 
of the measurements of different fictive temperatures and differences
 between two sets of measurements
 on the same series of samples. This value illustrates that light scattering
 data are in agreement with our SAXS measurements  for determination of 
the isothermal compressibility, but induce 
much larger errors.

The second result is the determination of the  relaxational 
compressibility using  a combination 
of the present SAXS data and previous Brillouin scattering
 measurements. This value is compared
 to data of the literature in column 2 of table \ref{tab:tab1}. 
Those data are calculated from isothermal compressibility in column
 1 and infinite frequency compressibility shown in column 3. 
All the values of $\chi_S^{\infty}$ are in relative agreement
 except that of Watanabe,\cite{WatanabeSaito2003} the only one  
not determined from Brillouin light scattering.
 Brillouin scattering is very accurate for the determination  of 
the infinite frequency compressibility 
because it relies on the position of Brillouin lines.  The differences 
in the values of infinite frequency 
compressibilities obtained from Brillouin scattering (less than $0.5 
\times 10^{-11}$ m s$^{2}$ kg$^{-1}$) 
can be assigned to either differences in samples (for example different 
OH contents\cite{TerkiLevelut1996}) 
or to differences in the density or refractive index used for the analysis 
(the index values used in the different papers differ from  up to 0.2\%). 
Except from our value, slightly lower, as the isothermal compressibility, 
all the  values of 
relaxational compressibilities are rather close, but that of Watanabe relies 
on  a very different
 value of infinite frequency compressibility. This point
will be discussed below.

The third result is  a new information brought out by our measurements: the combination of 
measurements below and above the glass transition to determine the 
isothermal compressibility as a function of fictive  temperature and temperature,
respectively.  Two different methods of analysis were 
used : we performed
in both cases  a full point by point analysis using
 equation (\ref{eqn:comp}) 
 in the supercooled liquid state on the one hand and using equation (\ref{eqn:Laberge}) 
in the glassy state on another hand. 
Previous studies consider either equation (\ref{eqn:comp}) above the
 glass transition, or equation (\ref{eqn:Laberge}), often applied
 at room temperature, assuming  that the 
infinite frequency inverse modulus and the isothermal 
compressibility\cite{SaitoIkushima1997,WatanabeSaito2003,BucaroDardy1974} 
 do not depend on temperature. A linear behavior is consequently obtained
 for  the intensity as 
a function of $T$ or $T_f$ and  the different contributions to 
compressibility are then
obtained by a global analysis using linear regressions. 
\cite{SaitoIkushima1997,WatanabeSaito2003} The error
    bars are  smaller than using a point by point analysis.
However, by using this point by point analysis we avoid any 
assumption on the temperature and 
fictive temperature
dependence of the different 
compressibility contributions.

For the fictive temperature dependence in the glass state, to our knowledge, the 
influence of thermal history on the isothermal 
and relaxational 
compressibilities have not been reported in the literature. 
 Indeed, Saito {\it et al}  have shown an influence 
of thermal history 
on the density fluctuations but their analysis rely on the assumption that
the compressibility do not depend on  the fictive temperature. Nevertheless,
 we found  a rather noticeable influence of heat treatment on the isothermal 
compressibility 
with a  variation of 8 \%  for $T_f$ varying between  $1373 \mbox{  K}$ and $1773\mbox{  K}$.
The changes are specially noticeable for the low $T_f$ 
 (when  $T_f$ is lowered from 1473 to $1373\mbox{ K}$). 
The error bar on $\chi_T^0$  being 
around 4\%, we considered this result as significant though  
at the limit of accuracy. 
This decrease with $T_f$  is a thoroughly new result.
Such dependence was suggested  by Le Parc et al\cite{LeParcChampagnon2001} when 
they discussed the fictive temperature dependence of  the Rayleigh intensity  
in silica at room
temperature. They suggest that  such a dependence could explain a non linear 
dependence 
of the  Rayleigh intensity, $I_{\mbox{\small Rayleigh}}$,  with $T_f$,
 because $I_{\mbox{\small Rayleigh}}\propto \chi_T^0(T_f) \times   T_f$.    
The variation of isothermal compressibility which occurs mostly  
at low fictive 
temperature  is 
 in qualitative agreement  with the variations of  $R_{LP}$ with 
$T_f$.\cite{LeParcChampagnon2002}

About the temperature dependence,  the fact that $\chi_T^0(T_f)$ decreases
 with increasing fictive temperature, is a very strong support for a decrease 
of the isothermal compressibility at least in the 1373-$1773\mbox{  K}$ range. 
We found almost the same temperature 
dependence of $\chi_T^0(T) $ in the supercooled liquid state, than that of 
$\chi_T^0(T_f)$ in the same temperature range (above 1530K), suggesting that 
$\chi_T^0(T)$ will follow the variation of $\chi_T^0(T_f)$ also below 
$1530\mbox{ K}$.
Several values of the literature for compressibilities correspond
 to determinations in the glass, $\chi_T^0(T_f)$, obtained using equation 
(\ref{eqn:Laberge}), around 1400K  
for most samples.\cite{WatanabeSaito2003,KrolLyons1986,Laberge1973} 
Most of them are above the value of $5.82 \pm 0.12
 \times 10^{-11}  \mbox{ m s}^2\mbox{  kg}^{-1}$  
found in the supercooled liquid. This also suggests  a decrease with
 increasing temperature of the isothermal 
compressibility. 
 Thus we want to emphasize  that the analysis 
performed in the glassy state and that performed in the liquid state are 
in very good agreement and they both suggest a decrease of the compressibility with 
temperature.  The small differences
observed  for example between $\chi_T^0(T_f)$ for $T_f=1373\mbox{  K}$ 
calculated using equation 
(\ref{eqn:Laberge})
 and the isothermal compressibility of the same sample in the liquid are 
compatible with 
a smooth decreases of $\chi_T^0$ with temperature in the range of 
the glass transition.
Such  smooth decreases of $\chi_T^0$ with temperature would be in 
 contrast with the data of Saito 
et al\cite{SaitoKakiuchida1997} which found an isothermal 
compressibility constant
 within glassy and supercooled state but with a discontinuity 
 and a strong increase
at the glass transition. We did not observe any strong discontinuity 
of the isothermal compressibility at the glass transition.

A variation of isothermal compressibility with both temperature and 
fictive temperature
induces that  the global analysis by linear fitting is a rather crude 
approximation. In order 
to evaluate the impact of the approximation made using this linear 
fitting,  we will now 
 use such an analysis on our SAXS data.
As we have measured both temperature and fictive temperature dependence 
of the SAXS intensity, we can  plot
 the intensity $I(q=0)$ either as a function of $T$ or $T_f$ and then use 
linear regressions or 
as a function of $T$ or $T_f$.

Using linear fits of $I(q=0)$ a function of $T$, we can first deduce a value of the isothermal
 compressibility $\chi_T^0$, supposed to be independent of temperature 
 using equation (\ref{eqn:comp}) for temperatures  above 
the glass transition. We obtain $\chi_T^0=5.52 \pm 0.22 \times 
10^{-11}\mbox{ m s}^2\mbox{  kg}^{-1}$. This value is very slightly lower, but 
close to  the values obtained by the point by point analysis. 
Below the glass transition, we can deduce 
$M_{\infty}^{-1}$ from the slope and $\chi_T^0-M_{\infty}^{-1}$ from 
the intercept
 of the fitting straight line. 
We found that a linear law can fit the data and 
 the slope yields  a high frequency inverse longitudinal modulus
$M_{\infty}^{-1}= 9.16 \pm 1.5 \times 10^{-12}\mbox{ m kg}^{-1}\mbox{ s}^2$.
 This value is  the average 
of the slopes obtained for each of the five samples
(with different $T_f$), the error  includes  differences between the 
determinations 
for different samples as well as the error on the fitting procedure.
This value is a bit lower   than that determined 
from our Brillouin scattering, from
$1.27 \times 10^{-11}\mbox{ m kg}^{-1}\mbox{ s}^2$   at room temperature to $1.12 
  \times 10^{-11}\mbox{ m kg}^{-1}\mbox{ s}^2$   at 1700 K, although it is in agreement
 with 
it, taking into account the error bar. The variation  with temperature  of 
  $M_{\infty}^{-1}$ determined using Brillouin measurements,  by 13\%,
 is below the error bar on the intercepts  determined from SAXS.
One can also verify that the value 
obtained using the slope of SAXS data as a function of temperature is
 much closer to $M_{\infty}^{-1}$ (Ref.\ 
~\onlinecite{Laberge1973,Schroder1973,BucaroDardy1974,KrolLyons1986}) than 
to $\chi_S^{\infty}$. These results consequently confirm
the validity of equation (\ref{eqn:Laberge}), 
where the temperature dependent
 term below $T_g$ is governed by $M_{\infty}^{-1}$  rather 
than $\chi_S^{\infty}$. The quantity $\chi_T^0-M_{\infty}^{-1}$,  
deduced from the 
intercept $T_f \left[ \chi_T^0(T_f)-M_{\infty}^{-1}(T_f)  \right] $
 of the same straight line used to determine $M_{\infty}^{-1}$, 
  slightly decreases
with increasing $T_f$, from  $5.15 \times 10^{-11}\mbox{ m kg}^{-1}\mbox{ s}^2$  
for $T_f= 1373$K to $4.7 \times 10^{-11}\mbox{ m kg}^{-1}\mbox{ s}^2$   for $T_f=1773$K. 
Those values are lower than the intercept
 obtained by Watanabe (table \ref{tab:tab1}) using the same measuring
 technique and the same analysis. Indeed, 
as stated above,  the intercept of  
Watanabe, corresponds, using our analysis with equation (\ref{eqn:Laberge}), to 
$\chi_T^0(T_f)-M_{\infty}^{-1}$  and not to  $\chi_T^r(T_f)$, as he assumed.
 One possible   reason for his higher value could 
rely on the measurements
themselves: the absolute SAXS intensity measured by Watanabe {\it et al.}
  are also slightly higher than the values usually reported in 
silica.\cite{LevelutGuinier1967,GerberHimmel,Weinberg1963,RenningerUhlmann1974,Bruening2005}
The values in the literature (except that of Ref.\ ~\onlinecite{WatanabeSaito2003}) 
are all around
18 and 25 electron units per molecule (4 and 5.5 $\times 10^{23}$ electron
unit per cm$^{3}$), 
 whereas the data of Watanabe et al are between
5.5 and 7.12 $\times 10^{23}$ electron unit per cm$^{3}$.
This difference could arise either
 from differences in the sample (different purities or different thermal 
histories) or
 from different  normalization procedure for  SAXS data (Watanabe normalized 
his SAXS data using the 
compressibility of quartz, Renninger used a Ludox solution, we used pure water).
 From the values of  $\chi_T^0(T_f)-M_{\infty}^{-1}$,  one can  calculated 
the isothermal compressibility, by adding
 $M_{\infty}^{-1}$ deduced from 
our Brillouin scattering data, or the relaxational compressibility by 
subtracting
 $\frac{4G_{\infty}}{M_{\infty}(3M_{\infty}-4G_{\infty})}$. One obtains 
then  $\chi_T^0$  and $\chi_S^r$ values which decrease with temperature from 
$\chi_T^0=6.3\times 10^{-11}\mbox{ m kg}^{-1}\mbox{ s}^2$   and $\chi_S^r=4 \times 
10^{-11}\mbox{ m kg}^{-1}\mbox{ s}^2$    for $T_f =1373$ K 
 to  $\chi_T^0=5.8 \times 10^{-11}\mbox{ m kg}^{-1}\mbox{ s}^2$   and $\chi_S^r=3.6 
\times 10^{-11}\mbox{ m kg}^{-1}\mbox{ s}^2$   for $T_f= 1773$ K.
 The results for the isothermal and relaxational compressibilities using
 this method are nearly identical to that of the point by point analysis.
In summary, the linear analysis as a function of temperature yield results 
in agreement with our point by point 
analysis, though it is less accurate.

Using 
our SAXS data measured for the different fictive temperatures, 
 we can also analyze the data
 using equation (\ref{eqn:Laberge}) as a function of fictive temperature 
for a given temperature of measurement. 
Now term 3 is constant and terms 1 and 2 vary with $T_f$.  If we performed this
 analysis as a function
 of fictive temperature  for different temperatures of measurement, 
we found a reasonable fit by a straight line, but the obtained
 $M_{\infty}^{-1}$ values are much too high (3 to 7 
times too high) and 
varies much faster than what is obtained using Brillouin scattering 
measurements. Indeed, this analysis  assumes that 
$\chi_T^0(T_f)-M_{\infty}^{-1}$ does not vary with $T_f$.
It  is equivalent
 to   a linear extrapolation down to $T_f=0$ of data measured for $T_f$ around 1300 
to 1800K. 
Consequently it induces large errors, specially because  there is not 
reason why the
 intensity as a function of $T_f$ should be linear. 
From our point by point analysis, we observe that $\chi_T^0$ vary with 
temperature and $T_f$, and so does $\chi_T^0(T_f)-M_{\infty}^{-1}$.
 There is consequently no reason to use a linear 
extrapolation of $I(q=0)$ with $T_f$.

Having shown the uncertainty of compressibility determination  using 
intensity measurements
 by light scattering and the limitations of a global analysis using linear
 fits of $I(q=0)$ determined by SAXS 
versus $T_f$,
 we come back to a comparison of our work with  other  small-angle measurements. 
Table \ref{tab:tab2} shows several determinations from 
 x-ray small-angle measurements and one from neutron small-angle x-ray 
scattering. In each line of  table \ref{tab:tab2} the first 
value of $\chi_T^0(T_f)$ is determined from the intensity 
measured at room temperature using equation (\ref{eqn:Laberge}) with 
$M_{\infty}^{-1}(300K)=1.27 \times 10^{-11} \mbox{ m s}^{2}\mbox{ kg}^{-1}$ 
in term 3 and $M_{\infty}^{-1}(T_f)\simeq 1.12 \times 10^{-11}
 \mbox{ m s}^{2}\mbox{ kg}^{-1}$ in term 1 and 2. 
The fictive temperature is either given by the authors either determined
 using viscosity
 data from Hetherington \cite{Hetherington1964} and a value of $10^{12}$ 
Pa. s for the viscosity 
at the glass transition 
($T_g=1473$K for infrared Vitreosil or 1400K for O. G. Vitreosil). 
The  values in next column is the value given by the authors themselves, 
when appropriate. For Ref.\ 
~\onlinecite{WatanabeSaito2003} and ~\onlinecite{Hulme} this values used $\chi_S^{\infty}$ in 
equation (\ref{eqn:Laberge}) 
instead of $M_{\infty}^{-1}$, which yields to an overestimate of the 
isothermal compressibility.
 Several determinations of SAXS intensity at room temperature using
 equation (\ref{eqn:Laberge}) 
are very close to our determination. 
\cite{LevelutGuinier1967,RenningerUhlmann1974} The compressibility 
value deduced from 
 Ref.\ ~\onlinecite{Hulme} is a bit higher but has a rather large error bar. 
 A few determinations are slightly higher than ours.
\cite{WatanabeSaito2003,Weinberg1963}  Differences in the studied 
silica samples (different impurity contents,
 different thermal history)  and  uncertainty 
on the glass transition temperature could be partly  responsible for the 
 differences. Differences up to 0.7 to
 $0.8 \times 10^{-11}\mbox{ m s}^{2}\mbox{ kg}^{-1}$  {\it i.e.} 10\% can be 
expected when the fictive temperature vary by  400 K. Differences
 about $0.15 \times 10^{-11}\mbox{ m s}^{2}\mbox{ kg}^{-1}$ can be expected when 
the OH content vary of 900 pmm.\cite{Bruening2005}

\begin{table}
\begin{tabular}{lll}\hline \hline
authors & $\chi _{T}(T_f)(\mbox{m s}^{2}\mbox{ kg}^{-1})$
&$\chi _{T}(\mbox{m s}^{2}\mbox{ kg}^{-1})$
 \\ \hline
& from Eq. (\ref{eqn:Laberge})& from authors\\
Watanabe($T_f=1230$K) \cite{WatanabeSaito2003}&
 $8.95 \times 10^{-11}$&$10.5\times 10^{-11}$\\
Watanabe($T_f=1670$K) \cite{WatanabeSaito2003}& 
$8.15 \times 10^{-11}$&$10.5\times 10^{-11}$\\
this work and  Ref.\ ~\onlinecite{LeParc2005} ($T_f=1373$K) & 
$6.16\pm 0.24\times 10^{-11}$ &\\
this work and Ref.\ ~\onlinecite{LeParc2005} ($T_f=1773$K) &
 $5.69 \pm 0.23 \times 10^{-11}$ &\\
Weinberg\cite{Weinberg1963}&$7.85 \times 10^{-11}$&
$7.8 \times 10^{-11}$\\
Levelut ($T_g=1400-1473 \mbox{ K}$)\cite{LevelutGuinier1967}
&$6.65-6.75\times 10^{-11}$&\\
Renninger ($T_g=1473$K)\cite{RenningerUhlmann1974}&$6.15\times 10^{-11}$&\\
Br\"uning GE-124 ($T_f=1535\mbox{ K}$)\cite{Bruening2005}&$5.85  \times 10^{-11}$& \\
Br\"uning Corning 7980 ($T_f=1303\mbox{ K}$)\cite{Bruening2005}&$6  \times 10^{-11}$& \\
Hulme (neutrons)\cite{Hulme}&$7.95 \pm0.7 \times 10^{-11}$&
$8.5 \pm0.7 \times 10^{-11}$\\ \hline \hline
\end{tabular}

\caption{\label{tab:tab2}Isothermal compressibility at $T_f$ deduced from 
SAXS intensity measurements $I(q=0)$, using equation (9), 
$M_{\infty}^{-1}(300K)=1.27 \times 10^{-11} \mbox{ m s}^{2}\mbox{ kg}^{-1}$ and 
$M_{\infty}^{-1}(T_f)=1.12 \times 10^{-11} \mbox{ m s}^{2}\mbox{ kg}^{-1}$. 
When the fictive temperature is not given in the paper, it is taken equal
 to the glass transition temperature 
(given by authors or deduced from  Ref.\ ~\onlinecite{Hetherington1964}, see text).}
\end{table}

\section{Conclusion}
To summarize, we want to emphasize 
that the values of $\chi_T(T_f)$ and $\chi_T(T)$ determined 
respectively from equation (\ref{eqn:Laberge})  and equation 
(\ref{eqn:comp})  present  temperature dependences which are
 in very good agreement (Fig. \ref{fig:betaT}).

In this paper, we reported   a complete determination 
of isothermal compressibility using SAXS measurements in several silica 
samples having different thermal histories. Using two different descriptions
 for the supercooled and the liquid state, 
we were able to determine the equilibrium isothermal compressibility $\chi_T^0$
as a function of temperature and as a function of fictive temperature, 
respectively. We obtained very comparable values from the two analysis, with 
the same temperature dependence. $\chi_T^0$
 is observed to decrease with fictive temperature in a non-linear manner from
6.2 to $5.7 \times 10^{-11} \mbox{ m s}^{2}\mbox{ kg}^{-1}$ in the
 range 1373 to 1773K. We also determined the fictive temperature dependence of the relaxational 
 compressibility, which is also decreasing  with fictive temperature.

In order to compare our data with previous determinations of the literature, 
we performed different  analysis on our data in order to evidence that differences 
can arise from the method of analysis.

We can also conclude from our full point by point analysis of data measured
 both as a function of temperature and fictive temperature, 
that Laberge description of the SAXS intensity in the glassy state is a
 very good  approximation which leads to results coherent with analysis 
in the supercooled liquid state.

To conclude, our values of isothermal compressibility  around 5.7 to $6.2 
 \times 10^{-11}\mbox{ m s}^2\mbox{  kg}^{-1}$,  are compatible with  most of the SAXS 
(or SANS) determinations, 
and not too far from the other previous determinations by light scattering , which 
are around $7 
 \times 10^{-11}\mbox{ m s}^2\mbox{  kg}^{-1}$, with a rather large error.

\section*{acknowledgements}
We wish to thank O. Geyamond, S. Arnaud, and B. Caillot (Laboratoire de Cristallographie, Grenoble) for technical assistance,
as well as J.-F. B\'erar (Laboratoire de Cristallographie, Grenoble) for assistance in using beamline BM02. We also thanks 
ESRF staff for operating the synchrotron radiation facilities.

\bibliography{levelutprb-art-rev}

\begin{thebibliography}{38}
\expandafter\ifx\csname natexlab\endcsname\relax\def\natexlab#1{#1}\fi
\expandafter\ifx\csname bibnamefont\endcsname\relax
  \def\bibnamefont#1{#1}\fi
\expandafter\ifx\csname bibfnamefont\endcsname\relax
  \def\bibfnamefont#1{#1}\fi
\expandafter\ifx\csname citenamefont\endcsname\relax
  \def\citenamefont#1{#1}\fi
\expandafter\ifx\csname url\endcsname\relax
  \def\url#1{\texttt{#1}}\fi
\expandafter\ifx\csname urlprefix\endcsname\relax\def\urlprefix{URL }\fi
\providecommand{\bibinfo}[2]{#2}
\providecommand{\eprint}[2][]{\url{#2}}

\bibitem[{\citenamefont{Tool and Eichlin}(1931)}]{Tool}
\bibinfo{author}{\bibfnamefont{A.~Q.} \bibnamefont{Tool}} \bibnamefont{and}
  \bibinfo{author}{\bibfnamefont{C.~G.} \bibnamefont{Eichlin}},
  \bibinfo{journal}{J. Am. Ceram. Soc.} \textbf{\bibinfo{volume}{14}},
  \bibinfo{pages}{276} (\bibinfo{year}{1931}).

\bibitem[{\citenamefont{Le~Parc and Levelut}(2005)}]{LeParc2005}
\bibinfo{author}{\bibfnamefont{R.}~\bibnamefont{Le~Parc}} \bibnamefont{and}
  \bibinfo{author}{\bibfnamefont{C.}~\bibnamefont{Levelut}}
  (\bibinfo{year}{2005}), \bibinfo{note}{to be submitted}.

\bibitem[{\citenamefont{Buchenau and Wischnewski}(2004)}]{Buchenau}
\bibinfo{author}{\bibfnamefont{U.}~\bibnamefont{Buchenau}} \bibnamefont{and}
  \bibinfo{author}{\bibfnamefont{A.}~\bibnamefont{Wischnewski}},
  \bibinfo{journal}{Physical Review B} \textbf{\bibinfo{volume}{70}},
  \bibinfo{pages}{092201} (\bibinfo{year}{2004}).

\bibitem[{\citenamefont{Weinberg}(1963)}]{Weinberg1963}
\bibinfo{author}{\bibfnamefont{D.~L.} \bibnamefont{Weinberg}},
  \bibinfo{journal}{Physics Letters} \textbf{\bibinfo{volume}{7}},
  \bibinfo{pages}{324} (\bibinfo{year}{1963}).

\bibitem[{\citenamefont{Wendorff and Fischer}(1973)}]{Wendorff1973}
\bibinfo{author}{\bibfnamefont{J.}~\bibnamefont{Wendorff}} \bibnamefont{and}
  \bibinfo{author}{\bibfnamefont{E.~W.} \bibnamefont{Fischer}},
  \bibinfo{journal}{Kolloid-Z. u. Z.Polymere} \textbf{\bibinfo{volume}{251}},
  \bibinfo{pages}{876} (\bibinfo{year}{1973}).

\bibitem[{\citenamefont{Roe and Curro}(1983)}]{Roe1983}
\bibinfo{author}{\bibfnamefont{R.-J.} \bibnamefont{Roe}} \bibnamefont{and}
  \bibinfo{author}{\bibfnamefont{J.~J.} \bibnamefont{Curro}},
  \bibinfo{journal}{Macromolecules} \textbf{\bibinfo{volume}{16}},
  \bibinfo{pages}{428} (\bibinfo{year}{1983}).

\bibitem[{\citenamefont{Curro and Roe}(1984)}]{Curro1984}
\bibinfo{author}{\bibfnamefont{J.~J.} \bibnamefont{Curro}} \bibnamefont{and}
  \bibinfo{author}{\bibfnamefont{R.-J.} \bibnamefont{Roe}},
  \bibinfo{journal}{Polymer} \textbf{\bibinfo{volume}{25}},
  \bibinfo{pages}{1424} (\bibinfo{year}{1984}).

\bibitem[{\citenamefont{Wiegand and Ruland}(1979)}]{Wiegand1979}
\bibinfo{author}{\bibfnamefont{W.}~\bibnamefont{Wiegand}} \bibnamefont{and}
  \bibinfo{author}{\bibfnamefont{W.}~\bibnamefont{Ruland}},
  \bibinfo{journal}{Progr. Colloid Polym. Sci.} \textbf{\bibinfo{volume}{66}},
  \bibinfo{pages}{355} (\bibinfo{year}{1979}).

\bibitem[{\citenamefont{Rathje and Ruland}(1976)}]{Rathje1976}
\bibinfo{author}{\bibfnamefont{J.}~\bibnamefont{Rathje}} \bibnamefont{and}
  \bibinfo{author}{\bibfnamefont{W.}~\bibnamefont{Ruland}},
  \bibinfo{journal}{Colloid Polym. Sci.} \textbf{\bibinfo{volume}{254}},
  \bibinfo{pages}{358} (\bibinfo{year}{1976}).

\bibitem[{\citenamefont{Laberge et~al.}(1973)\citenamefont{Laberge, Vasilescu,
  Montrose, and Macedo}}]{Laberge1973}
\bibinfo{author}{\bibfnamefont{N.~L.} \bibnamefont{Laberge}},
  \bibinfo{author}{\bibfnamefont{V.~V.} \bibnamefont{Vasilescu}},
  \bibinfo{author}{\bibfnamefont{C.~J.} \bibnamefont{Montrose}},
  \bibnamefont{and} \bibinfo{author}{\bibfnamefont{P.~B.}
  \bibnamefont{Macedo}}, \bibinfo{journal}{J. Amer. Ceram. Soc.}
  \textbf{\bibinfo{volume}{56}}, \bibinfo{pages}{506} (\bibinfo{year}{1973}).

\bibitem[{\citenamefont{Bucaro and Dardy}(1974)}]{BucaroDardy1974}
\bibinfo{author}{\bibfnamefont{J.}~\bibnamefont{Bucaro}} \bibnamefont{and}
  \bibinfo{author}{\bibfnamefont{H.~D.} \bibnamefont{Dardy}},
  \bibinfo{journal}{J. Applied Physics} \textbf{\bibinfo{volume}{46}},
  \bibinfo{pages}{5324} (\bibinfo{year}{1974}).

\bibitem[{\citenamefont{Krol et~al.}(1986)\citenamefont{Krol, Lyons, Brawer,
  and Kurkjian}}]{KrolLyons1986}
\bibinfo{author}{\bibfnamefont{D.~M.} \bibnamefont{Krol}},
  \bibinfo{author}{\bibfnamefont{K.~B.} \bibnamefont{Lyons}},
  \bibinfo{author}{\bibfnamefont{S.~A.} \bibnamefont{Brawer}},
  \bibnamefont{and} \bibinfo{author}{\bibfnamefont{C.~R.}
  \bibnamefont{Kurkjian}}, \bibinfo{journal}{Phys. Rev. B}
  \textbf{\bibinfo{volume}{33}}, \bibinfo{pages}{4196} (\bibinfo{year}{1986}).

\bibitem[{\citenamefont{Watanabe et~al.}(2003)\citenamefont{Watanabe, Saito,
  and Ikushima}}]{WatanabeSaito2003}
\bibinfo{author}{\bibfnamefont{T.}~\bibnamefont{Watanabe}},
  \bibinfo{author}{\bibfnamefont{K.}~\bibnamefont{Saito}}, \bibnamefont{and}
  \bibinfo{author}{\bibfnamefont{A.~J.} \bibnamefont{Ikushima}},
  \bibinfo{journal}{J. Applied Physics} \textbf{\bibinfo{volume}{94}},
  \bibinfo{pages}{4824} (\bibinfo{year}{2003}).

\bibitem[{\citenamefont{Polian et~al.}(2002)\citenamefont{Polian, Vo-Thang, and
  Richet}}]{Polian2002}
\bibinfo{author}{\bibfnamefont{A.}~\bibnamefont{Polian}},
  \bibinfo{author}{\bibfnamefont{D.}~\bibnamefont{Vo-Thang}}, \bibnamefont{and}
  \bibinfo{author}{\bibfnamefont{P.}~\bibnamefont{Richet}},
  \bibinfo{journal}{Europhysics Letters} \textbf{\bibinfo{volume}{57}},
  \bibinfo{pages}{375} (\bibinfo{year}{2002}).

\bibitem[{\citenamefont{Br{\"u}ckner}(1970)}]{Bruckner1970}
\bibinfo{author}{\bibfnamefont{R.}~\bibnamefont{Br{\"u}ckner}},
  \bibinfo{journal}{J. Non-Cryst. Solids} \textbf{\bibinfo{volume}{5}},
  \bibinfo{pages}{123} (\bibinfo{year}{1970}).

\bibitem[{\citenamefont{Saito et~al.}(1997)\citenamefont{Saito, Kakiuchida, and
  Ikushima}}]{SaitoKakiuchida1997}
\bibinfo{author}{\bibfnamefont{K.}~\bibnamefont{Saito}},
  \bibinfo{author}{\bibfnamefont{H.}~\bibnamefont{Kakiuchida}},
  \bibnamefont{and} \bibinfo{author}{\bibfnamefont{A.~J.}
  \bibnamefont{Ikushima}}, \bibinfo{journal}{J. Non-Crystallin Solids}
  \textbf{\bibinfo{volume}{222}}, \bibinfo{pages}{329} (\bibinfo{year}{1997}).

\bibitem[{\citenamefont{Saito and Ikushima}(1997)}]{SaitoIkushima1997}
\bibinfo{author}{\bibfnamefont{K.}~\bibnamefont{Saito}} \bibnamefont{and}
  \bibinfo{author}{\bibfnamefont{A.~J.} \bibnamefont{Ikushima}},
  \bibinfo{journal}{Appl. Phys. Lett.} \textbf{\bibinfo{volume}{70}},
  \bibinfo{pages}{3504} (\bibinfo{year}{1997}).

\bibitem[{\citenamefont{McSkimin}(1953)}]{McSkimin}
\bibinfo{author}{\bibfnamefont{H.~J.} \bibnamefont{McSkimin}},
  \bibinfo{journal}{J. Applied Phys.} \textbf{\bibinfo{volume}{24}},
  \bibinfo{pages}{988} (\bibinfo{year}{1953}).

\bibitem[{\citenamefont{Fine et~al.}(1954)\citenamefont{Fine, van Duyne, and
  Kenney}}]{Fine1954}
\bibinfo{author}{\bibfnamefont{M.~E.} \bibnamefont{Fine}},
  \bibinfo{author}{\bibfnamefont{H.}~\bibnamefont{van Duyne}},
  \bibnamefont{and} \bibinfo{author}{\bibfnamefont{N.~T.}
  \bibnamefont{Kenney}}, \bibinfo{journal}{J. appl. Phys.}
  \textbf{\bibinfo{volume}{25}}, \bibinfo{pages}{402} (\bibinfo{year}{1954}).

\bibitem[{\citenamefont{Spinner}(1956)}]{Spinner1956}
\bibinfo{author}{\bibfnamefont{S.}~\bibnamefont{Spinner}}, \bibinfo{journal}{J.
  Am. Ceram. Soc.} \textbf{\bibinfo{volume}{39}}, \bibinfo{pages}{113}
  (\bibinfo{year}{1956}).

\bibitem[{\citenamefont{Vukcevich}(1972)}]{Vukcevitch1972}
\bibinfo{author}{\bibfnamefont{M.~R.} \bibnamefont{Vukcevich}},
  \bibinfo{journal}{J. Non-Crystallin Solids} \textbf{\bibinfo{volume}{11}},
  \bibinfo{pages}{25} (\bibinfo{year}{1972}).

\bibitem[{\citenamefont{Champagnon et~al.}(2002)\citenamefont{Champagnon,
  Le~Parc, and Guenot}}]{ChampagnonLeParc2002}
\bibinfo{author}{\bibfnamefont{B.}~\bibnamefont{Champagnon}},
  \bibinfo{author}{\bibfnamefont{R.}~\bibnamefont{Le~Parc}}, \bibnamefont{and}
  \bibinfo{author}{\bibfnamefont{P.}~\bibnamefont{Guenot}},
  \bibinfo{journal}{Phil. Mag. B} \textbf{\bibinfo{volume}{82}}
  (\bibinfo{year}{2002}).

\bibitem[{\citenamefont{Champagnon et~al.}(2000)\citenamefont{Champagnon,
  Chemarin, Duval, and Le~Parc}}]{Champagnon2000}
\bibinfo{author}{\bibfnamefont{B.}~\bibnamefont{Champagnon}},
  \bibinfo{author}{\bibfnamefont{C.}~\bibnamefont{Chemarin}},
  \bibinfo{author}{\bibfnamefont{E.}~\bibnamefont{Duval}}, \bibnamefont{and}
  \bibinfo{author}{\bibfnamefont{R.}~\bibnamefont{Le~Parc}},
  \bibinfo{journal}{J. Non-Cryst. Solids} \textbf{\bibinfo{volume}{274}},
  \bibinfo{pages}{81} (\bibinfo{year}{2000}).

\bibitem[{\citenamefont{Le~Parc et~al.}(2001)\citenamefont{Le~Parc, Champagnon,
  Guenot, and Dubois}}]{LeParcChampagnon2001}
\bibinfo{author}{\bibfnamefont{R.}~\bibnamefont{Le~Parc}},
  \bibinfo{author}{\bibfnamefont{B.}~\bibnamefont{Champagnon}},
  \bibinfo{author}{\bibfnamefont{P.}~\bibnamefont{Guenot}}, \bibnamefont{and}
  \bibinfo{author}{\bibfnamefont{S.}~\bibnamefont{Dubois}},
  \bibinfo{journal}{J. Non-Cryst. Solids} \textbf{\bibinfo{volume}{293-295}},
  \bibinfo{pages}{366} (\bibinfo{year}{2001}).

\bibitem[{\citenamefont{Agarwal et~al.}(1975)\citenamefont{Agarwal, Davis, and
  Tomazawa}}]{Agarwal1955}
\bibinfo{author}{\bibfnamefont{A.}~\bibnamefont{Agarwal}},
  \bibinfo{author}{\bibfnamefont{K.~M.} \bibnamefont{Davis}}, \bibnamefont{and}
  \bibinfo{author}{\bibfnamefont{M.}~\bibnamefont{Tomazawa}},
  \bibinfo{journal}{J. Non-Cryst. Solids} \textbf{\bibinfo{volume}{220}}
  (\bibinfo{year}{1975}).

\bibitem[{\citenamefont{Geissberger and Galeener}(1983)}]{Galeener1983}
\bibinfo{author}{\bibfnamefont{A.~E.} \bibnamefont{Geissberger}} \bibnamefont{and}
  \bibinfo{author}{\bibfnamefont{F.~L.} \bibnamefont{Galeener}},
  \bibinfo{journal}{Phys. Rev. B} \textbf{\bibinfo{volume}{28}},
  \bibinfo{pages}{3266} (\bibinfo{year}{1983}).

\bibitem[{\citenamefont{Soldo et~al.}(1998)\citenamefont{Soldo, Hazemann,
  Aberdam, Inui, Tamura, Raoux, Pernot, Jal, and Dupuy-Philon}}]{Soldo1998}
\bibinfo{author}{\bibfnamefont{Y.}~\bibnamefont{Soldo}},
  \bibinfo{author}{\bibfnamefont{J.-L.} \bibnamefont{Hazemann}},
  \bibinfo{author}{\bibfnamefont{D.}~\bibnamefont{Aberdam}},
  \bibinfo{author}{\bibfnamefont{M.}~\bibnamefont{Inui}},
  \bibinfo{author}{\bibfnamefont{K.}~\bibnamefont{Tamura}},
  \bibinfo{author}{\bibfnamefont{D.}~\bibnamefont{Raoux}},
  \bibinfo{author}{\bibfnamefont{E.}~\bibnamefont{Pernot}},
  \bibinfo{author}{\bibfnamefont{J.-F.} \bibnamefont{Jal}}, \bibnamefont{and}
  \bibinfo{author}{\bibfnamefont{J.}~\bibnamefont{Dupuy-Philon}},
  \bibinfo{journal}{Phys. Rev. B.} \textbf{\bibinfo{volume}{57}},
  \bibinfo{pages}{258} (\bibinfo{year}{1998}).

\bibitem[{\citenamefont{Levelut and Guignier}(1967)}]{LevelutGuinier1967}
\bibinfo{author}{\bibfnamefont{A.~M.} \bibnamefont{Levelut}} \bibnamefont{and}
  \bibinfo{author}{\bibfnamefont{A.}~\bibnamefont{Guignier}},
  \bibinfo{journal}{Bull. Soc. Franc. Mineral. Crist.}
  \textbf{\bibinfo{volume}{90}} (\bibinfo{year}{1967}).

\bibitem[{\citenamefont{Levelut et~al.}(2002)\citenamefont{Levelut, Faivre,
  Le~Parc, Champagnon, Hazemann, David, Rochas, and Simon}}]{LevelutFaivre2002}
\bibinfo{author}{\bibfnamefont{C.}~\bibnamefont{Levelut}},
  \bibinfo{author}{\bibfnamefont{A.}~\bibnamefont{Faivre}},
  \bibinfo{author}{\bibfnamefont{R.}~\bibnamefont{Le~Parc}},
  \bibinfo{author}{\bibfnamefont{B.}~\bibnamefont{Champagnon}},
  \bibinfo{author}{\bibfnamefont{J.-L.} \bibnamefont{Hazemann}},
  \bibinfo{author}{\bibfnamefont{L.}~\bibnamefont{David}},
  \bibinfo{author}{\bibfnamefont{C.}~\bibnamefont{Rochas}}, \bibnamefont{and}
  \bibinfo{author}{\bibfnamefont{J.-P.} \bibnamefont{Simon}},
  \bibinfo{journal}{J. Non-Cristallin. Solids}
  \textbf{\bibinfo{volume}{307-310}}, \bibinfo{pages}{426}
  (\bibinfo{year}{2002}).

\bibitem[{\citenamefont{Champagnon}(2005)}]{Champagnonprivate}
\bibinfo{author}{\bibfnamefont{B.}~\bibnamefont{Champagnon}}
  (\bibinfo{year}{2005}), \bibinfo{note}{private communication}.

\bibitem[{\citenamefont{Terki et~al.}(1996)\citenamefont{Terki, Levelut,
  Boissier, and Pelous}}]{TerkiLevelut1996}
\bibinfo{author}{\bibfnamefont{F.}~\bibnamefont{Terki}},
  \bibinfo{author}{\bibfnamefont{C.}~\bibnamefont{Levelut}},
  \bibinfo{author}{\bibfnamefont{M.}~\bibnamefont{Boissier}}, \bibnamefont{and}
  \bibinfo{author}{\bibfnamefont{J.}~\bibnamefont{Pelous}},
  \bibinfo{journal}{Phys. Rev. B} \textbf{\bibinfo{volume}{53}},
  \bibinfo{pages}{2411} (\bibinfo{year}{1996}).

\bibitem[{\citenamefont{Le~Parc et~al.}(2002)\citenamefont{Le~Parc, Champagnon,
  David, Faivre, Levelut, Guenot, Hazemann, Rochas, and
  Simon}}]{LeParcChampagnon2002}
\bibinfo{author}{\bibfnamefont{R.}~\bibnamefont{Le~Parc}},
  \bibinfo{author}{\bibfnamefont{B.}~\bibnamefont{Champagnon}},
  \bibinfo{author}{\bibfnamefont{L.}~\bibnamefont{David}},
  \bibinfo{author}{\bibfnamefont{A.}~\bibnamefont{Faivre}},
  \bibinfo{author}{\bibfnamefont{C.}~\bibnamefont{Levelut}},
  \bibinfo{author}{\bibfnamefont{P.}~\bibnamefont{Guenot}},
  \bibinfo{author}{\bibfnamefont{J.-L.} \bibnamefont{Hazemann}},
  \bibinfo{author}{\bibfnamefont{C.}~\bibnamefont{Rochas}}, \bibnamefont{and}
  \bibinfo{author}{\bibfnamefont{J.-P.} \bibnamefont{Simon}},
  \bibinfo{journal}{Philos. Mag. B} \textbf{\bibinfo{volume}{82}},
  \bibinfo{pages}{431} (\bibinfo{year}{2002}).

\bibitem[{\citenamefont{Shroeder et~al.}(1973)\citenamefont{Shroeder, Mohr,
  Macedo, and Montrose}}]{Schroder1973}
\bibinfo{author}{\bibfnamefont{J.}~\bibnamefont{Shroeder}},
  \bibinfo{author}{\bibfnamefont{R.}~\bibnamefont{Mohr}},
  \bibinfo{author}{\bibfnamefont{P.~B.} \bibnamefont{Macedo}},
  \bibnamefont{and} \bibinfo{author}{\bibfnamefont{C.~J.}
  \bibnamefont{Montrose}}, \bibinfo{journal}{J. Amer. Ceram. Soc.}
  \textbf{\bibinfo{volume}{56}}, \bibinfo{pages}{510} (\bibinfo{year}{1973}).

\bibitem[{\citenamefont{Gerber and Himmel}(1986)}]{GerberHimmel}
\bibinfo{author}{\bibfnamefont{T.}~\bibnamefont{Gerber}} \bibnamefont{and}
  \bibinfo{author}{\bibfnamefont{B.}~\bibnamefont{Himmel}},
  \bibinfo{journal}{J. Non-Cristallin. Solids} \textbf{\bibinfo{volume}{83}},
  \bibinfo{pages}{324} (\bibinfo{year}{1986}).

\bibitem[{\citenamefont{Renninger and Uhlmann}(1974)}]{RenningerUhlmann1974}
\bibinfo{author}{\bibfnamefont{A.~L.} \bibnamefont{Renninger}}
  \bibnamefont{and} \bibinfo{author}{\bibfnamefont{D.~R.}
  \bibnamefont{Uhlmann}}, \bibinfo{journal}{J. Non-Cryst. Solids}
  \textbf{\bibinfo{volume}{16}}, \bibinfo{pages}{325} (\bibinfo{year}{1974}).

\bibitem[{\citenamefont{Br{\"u}ning et~al.}(2005)\citenamefont{Br{\"u}ning,
  Levelut, Faivre, Le~Parc, Simon, Bley, and Hazemann}}]{Bruening2005}
\bibinfo{author}{\bibfnamefont{R.}~\bibnamefont{Br{\"u}ning}},
  \bibinfo{author}{\bibfnamefont{C.}~\bibnamefont{Levelut}},
  \bibinfo{author}{\bibfnamefont{A.}~\bibnamefont{Faivre}},
  \bibinfo{author}{\bibfnamefont{R.}~\bibnamefont{Le~Parc}},
  \bibinfo{author}{\bibfnamefont{J.-P.} \bibnamefont{Simon}},
  \bibinfo{author}{\bibfnamefont{F.}~\bibnamefont{Bley}}, \bibnamefont{and}
  \bibinfo{author}{\bibfnamefont{J.-L.} \bibnamefont{Hazemann}},
  \bibinfo{journal}{Europhysics Letters} \textbf{\bibinfo{volume}{70}},
  \bibinfo{pages}{211} (\bibinfo{year}{2005}).

\bibitem[{\citenamefont{Hetherington et~al.}(1964)\citenamefont{Hetherington,
  Jack, and Kennedy}}]{Hetherington1964}
\bibinfo{author}{\bibfnamefont{G.}~\bibnamefont{Hetherington}},
  \bibinfo{author}{\bibfnamefont{K.~H.} \bibnamefont{Jack}}, \bibnamefont{and}
  \bibinfo{author}{\bibfnamefont{J.~C.} \bibnamefont{Kennedy}},
  \bibinfo{journal}{Phys. Chem. Glasses} \textbf{\bibinfo{volume}{5}},
  \bibinfo{pages}{130} (\bibinfo{year}{1964}).

\bibitem[{\citenamefont{Hulme}(1991)}]{Hulme}
\bibinfo{author}{\bibfnamefont{R.}~\bibnamefont{Hulme}}, Ph.D. thesis,
  \bibinfo{school}{University of Reading (U.K)} (\bibinfo{year}{1991}).

\end{thebibliography}
\bibliographystyle{apsrev}

\end{document}